\documentclass[]{spie}  %>>> use for US letter paper
\usepackage{ifthen}
\usepackage{xcolor}
\usepackage[]{graphicx}
\usepackage{amssymb}
\usepackage{floatrow}
\newfloatcommand{capbtabbox}{table}[][\FBwidth]

\def\gt{\ensuremath{>}}
\def\sgras{SgrA*}

\title{Making SPIFFI SPIFFIER: Upgrade of the SPIFFI instrument for use in ERIS and performance analysis from re-commissioning.} 

\def\MPE{a}
\def\ESO{c}
\def\weisz{b}

\author{
\mbox{E.~M.~George\supit{\MPE},}
\mbox{D.~Gr{\"a}ff\supit{\MPE},}
\mbox{H.~Feuchtgruber\supit{\MPE},}
\mbox{M.~Hartl\supit{\MPE},}
\mbox{F.~Eisenhauer\supit{\MPE},}
\mbox{A.~Buron\supit{\MPE},}
\mbox{R.~Davies\supit{\MPE},}
\mbox{R.~Genzel\supit{\MPE},}
\mbox{H.~Huber\supit{\MPE},}
\mbox{C.~Rau\supit{\MPE},}
\mbox{M.~Plattner\supit{\MPE},}
\mbox{E.~Wiezorrek\supit{\MPE},}
\mbox{H.~Weisz\supit{\weisz},}
\mbox{P.~Amico\supit{\ESO},}
\mbox{A.~Glindeman\supit{\ESO},}
\mbox{G.~Hau\supit{\ESO},}
\mbox{H.~Kuntschner\supit{\ESO},}
\mbox{A.~Modigliani\supit{\ESO}}
\skiplinehalf
\footnotesize{
\supit{a}Max Planck Institute f{\"u}r Extraterrestrische Physik, Giessenbachstasse 1, 87458 Garching, Germany \\
\supit{b}WEISZ Ing.-Bureau f{\"u}r den Maschinenbau, Germany\\ 
\supit{c}European Southern Observatory, Germany
}
}

\authorinfo{Further author information: (Send correspondence to E.M. George) \\ E-mail: lizinvt@mpe.mpg.de, Telephone: +49 89 30000 3283}
\begin{document} 
\maketitle

%%%%%%%%%%%%%%%%%%%%%%%%%%%%%%%%%%%%%%%%%%%%%%%%%%%%%%%%%%%%% 
\begin{abstract}

SPIFFI is an AO-fed integral field spectrograph operating as part of SINFONI on the VLT, which will be upgraded and reused as SPIFFIER in the new VLT instrument ERIS. In January 2016, we used new technology developments to perform an early upgrade to optical subsystems in the SPIFFI instrument so ongoing scientific programs can make use of enhanced performance before ERIS arrives in 2020. We report on the upgraded components and the performance of SPIFFI after the upgrade, including gains in throughput and spatial and spectral resolution.  We show results from re-commissioning, highlighting the potential for scientific programs to use the capabilities of the upgraded SPIFFI. Finally, we discuss the additional upgrades for SPIFFIER which will be implemented before it is integrated into ERIS.

\end{abstract}

%>>>> Include a list of keywords after the abstract 


%%%%%%%%%%%%%%%%%%%%%%%%%%%%%%%%%%%%%%%%%%%%%%%%%%%%%%%%%%%%%
\section{INTRODUCTION}
\label{sec:intro}  % \label{} allows reference to this section
SPIFFI\cite{iserlohe04} (SPectrometer for Infrared Faint Field Imaging) is an adaptive optics fed near-infrared imaging spectrograph that has been operational as a facility instrument for the ESO Very Large Telescope (VLT) as part of the SINFONI\cite{eisenhauer03} project since 2005. The instrument has been scientifically productive in several areas over its 11-year lifetime, and is currently in use for several high-profile scientific programs. An upgraded version of the instrument will be included in the new VLT Adaptive Optics (AO) instrument ERIS\cite{amico12, kuntschner14} (Enhanced Resolution Imager and Spectrometer) as the Integral Field Unit (IFU) subsystem SPIFFIER (SPectrometer for Infrared Faint Field Imaging Enhanced Resolution).  

We describe the planned changes for the upgraded instrument, which include upgrades to almost all subsystems in the instrument. These consist of new motors and electronics, new pre-optics, new filters, new spectrometer collimator mirrors, at least one new diffraction grating, and a new detector. Technology developments in many areas during the last 15 years enable us to upgrade this successful instrument to be competitive as a second-generation instrument on the VLT and make good use of the new 4 laser guide star adaptive optics facility (4LGSF) on VLT Unit Telescope 4 (UT4). These developments include modern multi-layer coatings for lenses and filters, more precise diamond turning processes in mirror manufacturing, immersion grating development, and quantum efficiency improvements in new generations of infrared detectors. We discuss the technology developments that enable us to make the SPIFFI instrument SPIFFIER, as well as the scientific performance gains expected from these upgrades.

In January 2016, we performed an early upgrade to some of the subsystems in the SPIFFI instrument so that ongoing scientific programs can make use of enhanced performance before ERIS arrives in 2020. Section \ref{sec:early} discusses the early upgrade, in which the optical components making up the pre-optics subsystem, bandpass filters, and spectrometer collimator mirrors were exchanged, resulting throughput and spectral resolution gains. We report the performance of SPIFFI after this early upgrade in section \ref{sec:performance}. We show results from the re-commissioning, highlighting the scientific potential for on-going and new science programs using the new capabilities of the upgraded instrument. 

The final upgrades to SPIFFIER will occur shortly before integration into ERIS. In section \ref{sec:eris}, we discuss the changes to the ERIS design since the publication of Kuntschner et al 2014\cite{kuntschner14}, and the further upgrades to be performed to SPIFFIER. We conclude in section \ref{sec:conclusion}.

%%%%%%%%%%%%%%%%%%%%%%%%%%%%%%%%%%%%%%%%%%%%%%%%%%%%%%%%%%%%%
\section{The SPIFFI instrument}
\label{sec:spiffi}

SPIFFI is an integral field spectrograph that allows a 3D view of astronomical objects (one wavelength and two spatial dimensions). Figure \ref{fig:spiffi} shows the layout of the instrument. SPIFFI is currently being used with the adaptive optics system MACAO\cite{bonnet03} as part of SINFONI, which provides a nearly diffraction-limited beam from the 8-meter UT4 telescope. The design and performance of SPIFFI, MACAO, and SINFONI are documented elsewhere,\cite{eisenhauer03, iserlohe04, bonnet04} however, as we will discuss upgrades and modifications to the SPIFFI optics, a short description and instrument layout are provided here.

\subsection{Optical Layout}
\label{sec:layout}

\begin{figure}[htbp!]
\begin{center}
\resizebox{0.5\textwidth}{!}{
\includegraphics[width=1.0\textwidth]{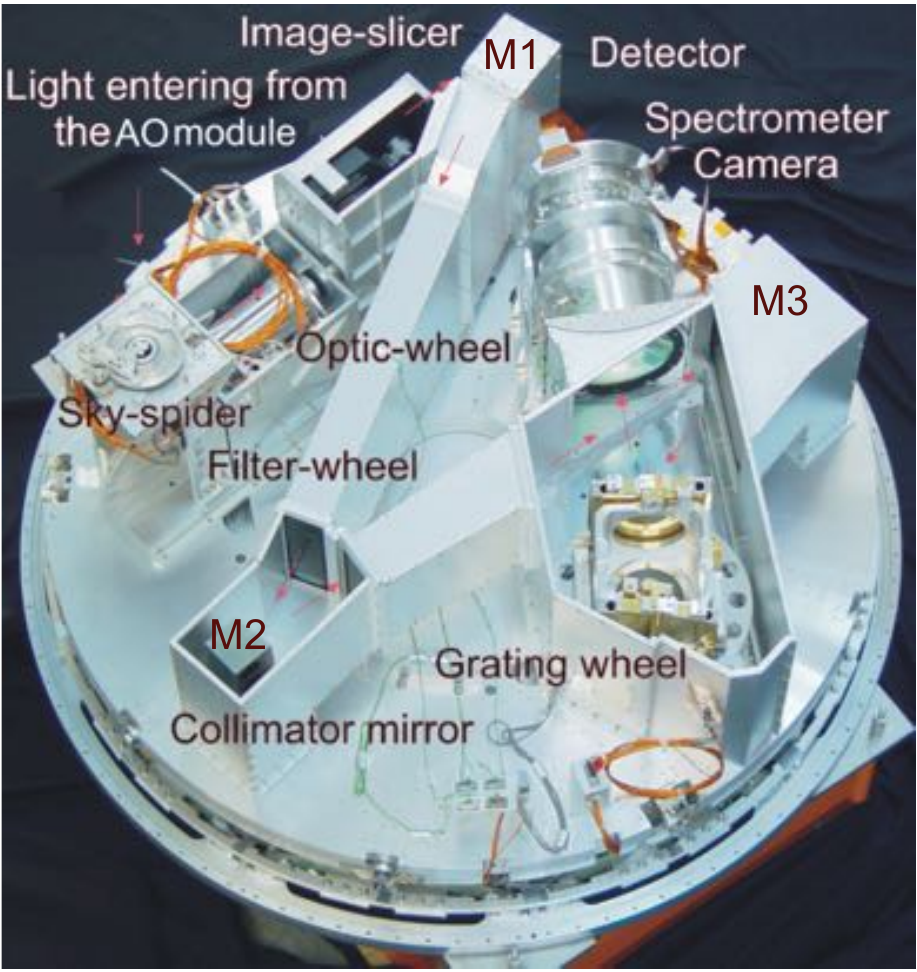}
}
\caption{Layout of SPIFFI with the light path shown in red arrows. See text in section \ref{sec:spiffi} for an explanation of each component. The spectrometer camera shown is the original 1k camera, it has been replaced in 2005 with a new 2K camera (See figure 8 of Kroes et al 2008\cite{kroes08} for a picture of the new 2K camera). }
\label{fig:spiffi}
\end{center}
\end{figure}

After passing through the AO module, the IR light enters the SPIFFI cryostat through a dichroic entrance window from above, and passes through several cold baffles on the radiation shields before entering the sky spider, which sits in the entrance focal plane.\footnote{While the sky spider was initially intended to allow simultaneous sky observations on fields up to 45" away negating the need for separate sky observations, the spatially varying and complex instrumental line profile of the instrument prevents its use in observational programs.} The light then enters the pre-optics collimator, where after a 45$^o$ mirror it passes through a filter wheel that provides filters for J, H, K, and H+K bands. The light then passes a cold lyot stop, which contains a central obscuration to mask thermal emission from the secondary mirror. Next the light enters the pre-optics wheel, which contains four cameras. Three of these provide different pixel scales on-sky of 25, 100, and 250 milliarcseconds (mas), while the final camera is a pupil-imaging camera used for alignment between SPIFFI and MACAO. The pre-optics cameras image the object plane onto the image slicer, which slices the 2D field into 32 strips and re-arranges them into a pseudo-slit (see figure 2 of Iserlohe et al 2004\cite{iserlohe04} for a layout of the image slicer). 

The light is now in the spectrograph portion of the instrument. The light is collimated with one spherical and two off-axis oblate elliptical mirrors, which have been produced via diamond-turning. The collimated light falls on one of four diffraction gratings located in the pupil position, which reflect and disperse the light to cover the full J, H, K or H+K bands with a design resolution of ~4500 for the single bands and ~2200 for the H+K band grating. The light then enters a f/2.8 camera developed by ASTRON/NOVA\cite{kroes08}, which was installed in 2005. Unfortunately, we do not have an image of the instrument with the new camera, as it was installed in-situ after all of the stiffening structure was installed around the optics, so the 1K camera is shown in figure \ref{fig:spiffi}. Figure 8 of Kroes et al 2008\cite{kroes08} shows the 2K camera and Hawaii 2RG detector housing. The spectrum is then imaged onto a HAWAII 2RG detector from Rockwell, providing Nyquist sampling of the spectrum in a single integration. After wavelength calibration, the 2D detector image of the pseudo-slit spectrum can be reassembled into a 64x32 pixel image of the object plane, where each pixel contains the spectrum of a full wavelength band.

%%%%%%%%%%%%%%%%%%%%%%%%%%%%%%%%%%%%%%%%%%%%%%%%%%%%%%%%%%%%%
\subsection{Science goals}
\label{sec:science}

SPIFFI as part of SINFONI has been very scientifically productive over the first 11 years of its use. We discuss briefly a few science programs that have used SINFONI over the years to highlight the scientific potential of the newly upgraded instrument. 

SPIFFI was designed specifically for studying the Galactic Center, and indeed some of the first science results from the SINFONI instrument were the discovery of young B stars in the central light-month, named the ``S-stars", orbiting the central black hole (\sgras), and characterisation of the spectra of infrared flares of \sgras.\cite{eisenhauer05,gillessen06} The orbits of the S-stars produced over many years of monitoring have been used to measure the mass of and distance to \sgras.\cite{gillessen09} The long-term monitoring of the orbit and upcoming peri-center passage of the star S2 provides an opportunity to test the theory of general relativity in the strong-field limit.\cite{gillessen09a,genzel10} The long-term monitoring of \sgras\ with SINFONI also produced some surprise discoveries--such as the 2012 discovery of a gas cloud, named G2, falling toward the central black hole.\cite{gillessen12} Since that time, the nature of G2 has been the subject of much debate in the Galactic Center community and its passage through pericenter has been studied closely by several groups using SINFONI,\cite{gillessen13, gillessen13a,gillessen14,pfuhl15} as well as triggering follow-up studies in many other instruments covering a large wavelength range.\cite{phifer13,haggard15}

Another key area of study using the SINFONI instrument as been in the area of high-redshift galaxy formation and evolution.\cite{genzel06} 
The SINS survey was the first and largest survey of spatially resolved gas kinematics, star formation distribution, and ionized gas properties of star-forming galaxies at z $\sim$1-3, with 110 sources observed in seeing-limited mode and 35 of them followed up with adaptive optics reaching angular resolutions down to $\sim$ 0.1" \cite{forsterschreiber09,genzel14}.  
This, and other SINFONI surveys at z $\gtrsim$ 1 (e.g. MASSIV\cite{epinat09, epinat12}) notably provided some of the key empirical evidence that smoother accretion via cold gas streams along the cosmic web and minor mergers dominate the mass assembly of massive galaxies a few billion years after the Big Bang, with a majority of rotating -- yet turbulent -- gas-rich disks.  
SINFONI+AO observations also resulted in the first detection of the roots of powerful gas outflows originating from intensely star-forming complexes in the disks of high redshift galaxies as well as AGN-driven winds from the nuclear regions in the most massive ones.\cite{newman12, forsterschreiber14}.  
SINFONI also enabled measurements of metallicity gradients in high-redshift galaxies, out to z $\sim$ 3 where some isolated galaxies displayed an unexpected decrease in oxygen abundance towards the center that could signal significant dilution from rapidly inflowing primordial gas.\cite{cresci10}

Both of these science cases will benefit from the upgraded instrument, for example from improved instrumental line profiles and increased throughput.
Additionally, the galactic center science case requires very good AO performance, which while currently quite good, will be further improved in ERIS.
The studies of galaxy formation and evolution using measurements of H$\alpha$ in the redshift range around z$\sim$1 will benefit from increased throughput in J-band, and additionally from improved spectral resolution, which is currently limited to $\sim$2000 in J-band.

%%%%%%%%%%%%%%%%%%%%%%%%%%%%%%%%%%%%%%%%%%%%%%%%%%%%%%%%%%%%%
\section{Early upgrade}
\label{sec:early}

In January 2016, we upgraded some components of SPIFFI so that ongoing science projects could take advantage of advanced performance before ERIS arrives. Here we describe the components upgraded.

%%%%%%%%%%%%%%%%%%%%%%%%%%%%%%%%%%%%%%%%%%%%%%%%%%%%%%%%%%%%%
\subsection{Pre-optics}
\label{sec:preoptics}

The pre-optics system consists of a pre-optics collimator, a filter wheel with selectable filters, a cold lyot stop with centeral obscuration, and a pre-optics wheel with four cameras providing imaging on-sky with a pixel scale of 25, 100, or 250 mas, as well as a pupil imaging camera used for alignment. An optical description for the pre-optics can be found in section 2.2 of Eisenhauer et al 2000\cite{eisenhauer00}. In the 13 years since the original optics were integrated, the pre-optics system has shown some wear. In particular, in an intervention in 2013, it was found that at least one BaF$_2$ lens in the 25 mas camera had damaged and peeling AR coatings (see figure 7d of Eisenhauer et al 2003\cite{eisenhauer03} for an example of damaged coatings). This lens was polished to remove the damaged coating, leaving this lens with no AR coating. Since many lenses are not visible inside of the lens tubes, it was not possible to assess how many other lenses could be affected by coating damage, and the decision was made to replace the entire pre-optics system.  

\begin{figure}[htbp!]
\begin{center}
\resizebox{0.7\textwidth}{!}{
\includegraphics[width=1.0\textwidth,angle=90]{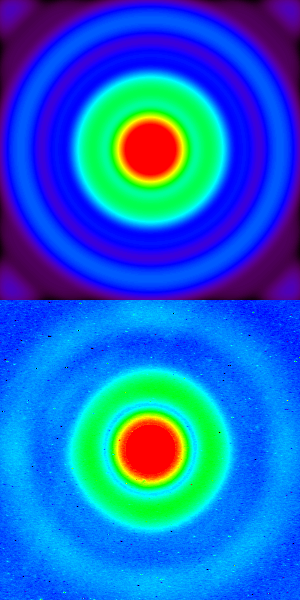}
}
\caption{Image of theoretical ({\it left}) and measured ({\it right}) PSF of the new pre-optics in the 25 mas camera in K-band as measured during lab tests. The theoretical PSF does not take into account the four spiders supporting the central obscuration in the lyot stop. The effects of these spiders can be seen in the measured PSF. }
\label{fig:preopticspsf}
\end{center}
\end{figure}

For the early upgrade, we manufactured a completely new pre-optics system, including all wheels and motors, and tested it in a dedicated cryogenic test facility with a Hawaii1K detector at the location of the image slicer. This allowed a higher spatial sampling of the PSF than is possible in the SPIFFI instrument, and more importantly, allowed us to directly compare our measurements of the Strehl Ratio to those taken on the original pre-optics without worrying about the effects of different sampling of the PSF.\cite{roberts04} Figure \ref{fig:preopticspsf} shows an example of a measured PSF of the pre-optics system in our test facility, compared with a theoretically calculated PSF. Table \ref{tab:preoptics} tabulates the measured Strehl ratios, fractional encircled energies within a single slitlet on the small slicer (300 microns in the image plane), and the improvement in Strehl ratio as compared to the old pre-optics (defined as New SR - Old SR). The Strehl ratio was calculated by normalizing the flux within a radius of $7\lambda/D$ to unity, and compared to a theoretically calculated PSF which takes into account the finite wavelength range and central obscuration, but not the four support spiders for the central obscuration. The error in the SR measurement is of order 5-10\%, resulting from our imperfect calculation of the theoretical PSF and noise in the measurements.

\begin{table}[htb]
\begin{center}
\begin{tabular}{ccccc}
\hline
Band  & Pixel scale & Strehl Ratio & Encircled Energy fraction& Increase in SR\\
 & [mas] & [\%] &[\%] & [$\Delta$\%]\\
\hline
\hline
J & 25&91 &90 & -\\
 & 100&85 &91 &5 \\
 & 250&106 &99 & 51\\
 \hline
K & 25&105 &101 & -\\
 & 100&97 &100 & -2\\
 & 250&101 &99 & 25\\
 \hline
\end{tabular}
\vspace{0.05in}
\caption{Table of measured Strehl Ratio, fractional encircled energy within a slitlet (measured/theoretical), and improvement in the SR between the old and new pre-optics (defined as New SR - Old SR) as measured in pre-deployment laboratory tests. Measurements were only taken in J and K bands, and the old 25mas pre-optics measurements are not available. The error on the SR measurement is of order 5-10\%. }
\label{tab:preoptics}
\end{center}
\end{table}

We find a significant improvement in SR for the 250 mas pixel scale in both J and K bands. In the 100mas pixel scale, the measured strehl ratios are identical within the error bars. We were unable to assess the improvement in the 25 mas pixel scale, as the original lab measurements using the old pre-optics of the 25 mas pixel scale were unavailable, however, we measured very high SRs in J and K bands in the 25 mas pixel scale. Finally, the modern AR coatings applied to the new lenses have slightly higher transmission than the old AR coatings ($\gt98.5\%$ vs. $\gt98\%$ over all bands), providing a modest improvement in throughput, given that there are 8 or 10 lens surfaces in the pre-optics system depending on the camera selected. Overall, the new pre-optics provides a performance enhancement in throughput and SR over the old pre-optics, and should provide optimal performance throughout the lifetime of the ERIS instrument. Section \ref{sec:image} shows the SR obtained on-sky with the new pre-optics.

%%%%%%%%%%%%%%%%%%%%%%%%%%%%%%%%%%%%%%%%%%%%%%%%%%%%%%%%%%%%%
\subsection{Filters}
\label{sec:filters}

The pre-optics system contains a filter wheel with selectable band-pass filters for J, H, K, and H+K bands. The filters are designed to have the maximum transmission in band, and suppress transmission of wavelengths outside of the band that would result in transmission of unwanted grating orders to the detector. The original filters in SPIFFI were manufactured in the year 2000, when multi-layer coatings with high out-of-band suppression were limited in the number of layers that could be deposited. Despite being very high quality for the time, the in-band transmission shows many interference fringes resulting in oscillating transmission across the band.

\begin{figure}[htbp!]
\begin{center}
\resizebox{1.0\textwidth}{!}{
\includegraphics[trim={2cm 0 1.8cm 0},clip=true,width=1.0\textwidth]{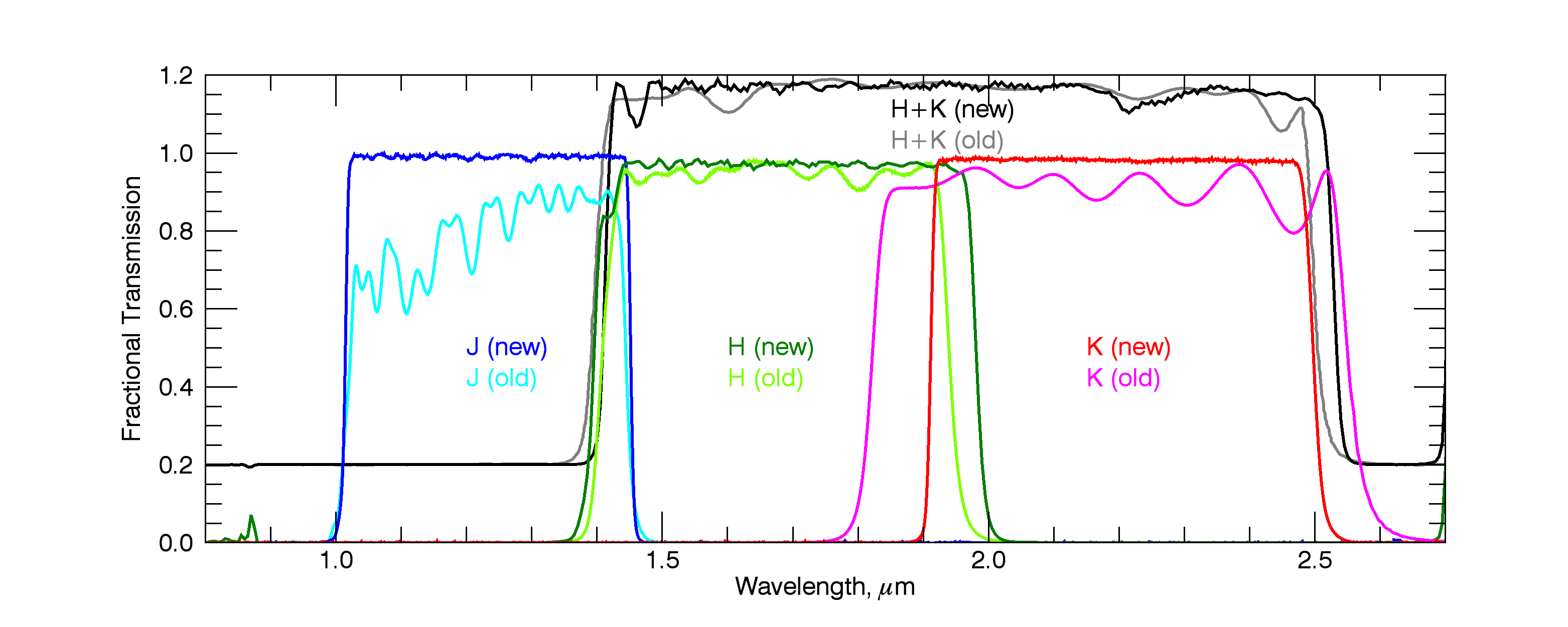}
}
\caption{Plot of the new filter transmission curves vs. the old filter curves. The H+K filters have been offset by +0.2 units for plot readability. The new filters have many more layers, and hence much smaller (but higher frequency) oscillations in the transmission, as well as an overall higher throughput. The out-of-band suppression between 0.8 and 2.7 $\mu$m is excellent. Starting with the January 2016 upgrade, the default filters in use are J (new), H (new), K (new), and H+K (old).}
\label{fig:filters}
\end{center}
\end{figure}

For the early upgrade we obtained new J and K band filters from IRIDIAN, which use a large number of layers to achieve extremely high and uniform transmission in-band of $\gt99\%$ in J-band and $\gt98\%$ in K-band, with excellent out-of-band blocking. We also obtained new H and H+K band filters from Laser Components on a short time-scale to serve as spares. The H-band filter shows significantly better (and more uniform) transmission than the old one ($\gt97\%$ in band) and thus was also installed during the upgrade. The new H+K band filter did not show significantly better transmission than the old one due to the substrate on which it was manufactured, and was therefore left as a spare. Figure \ref{fig:filters} shows the filter transmission curves of the new and old J, H, K, and H+K band filters. The transmission was improved by up to 35$\%$ (short wavelengths of J-band) by replacing the filters. Section \ref{sec:throughput} shows a measurement of overall instrument throughput improvement on-sky.

%%%%%%%%%%%%%%%%%%%%%%%%%%%%%%%%%%%%%%%%%%%%%%%%%%%%%%%%%%%%%
\subsection{Collimator mirrors}
\label{sec:mirrors}

\begin{figure}[htbp!]
\begin{center}
\resizebox{1.0\textwidth}{!}{
\includegraphics[height=6cm]{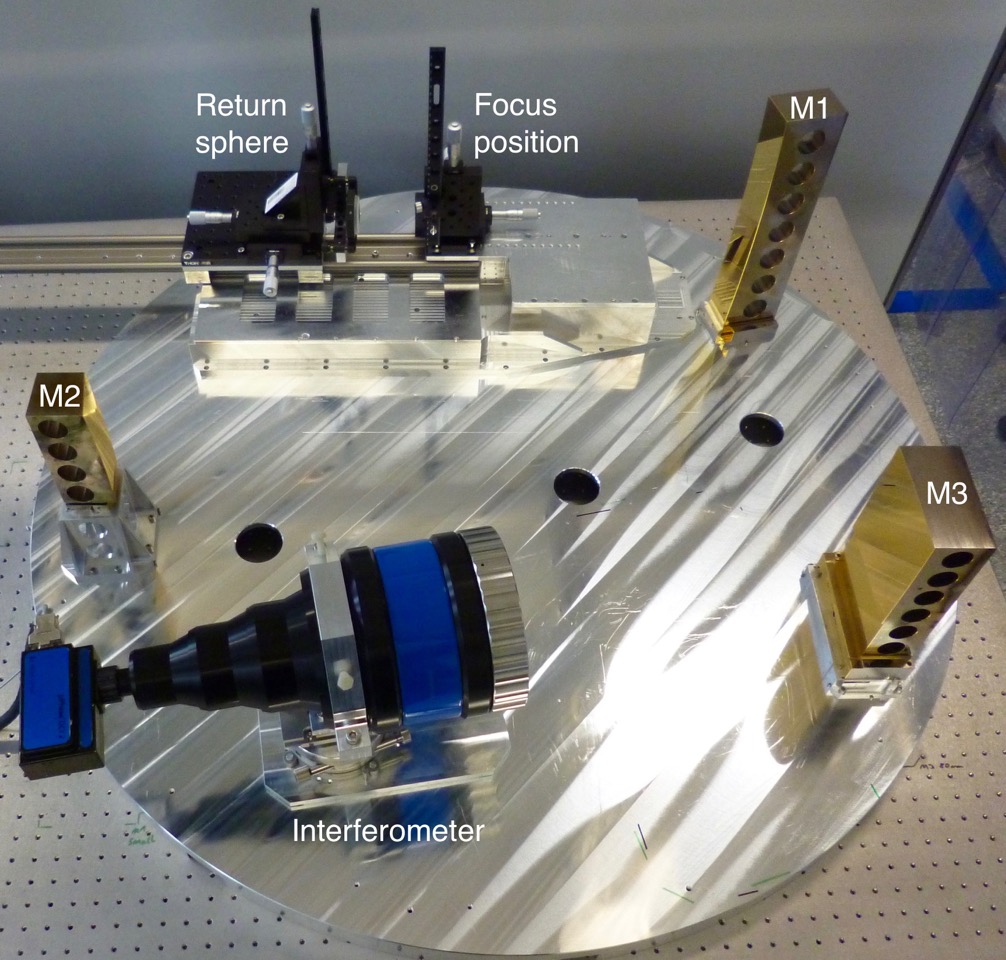}
\includegraphics[height=6cm, trim={0 0 2.5cm 0}, clip=true]{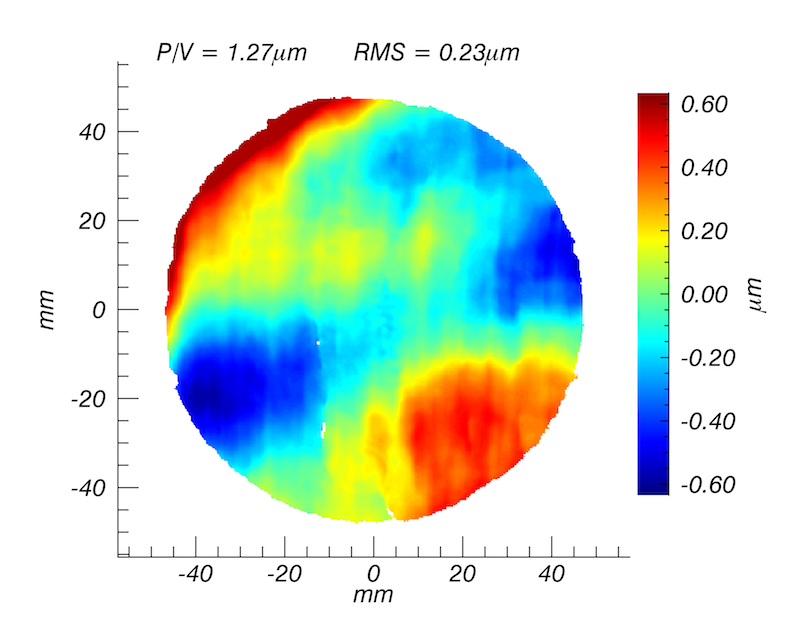}
\includegraphics[height=6cm, trim={2.0cm 0 0.5cm  0}, clip=true]{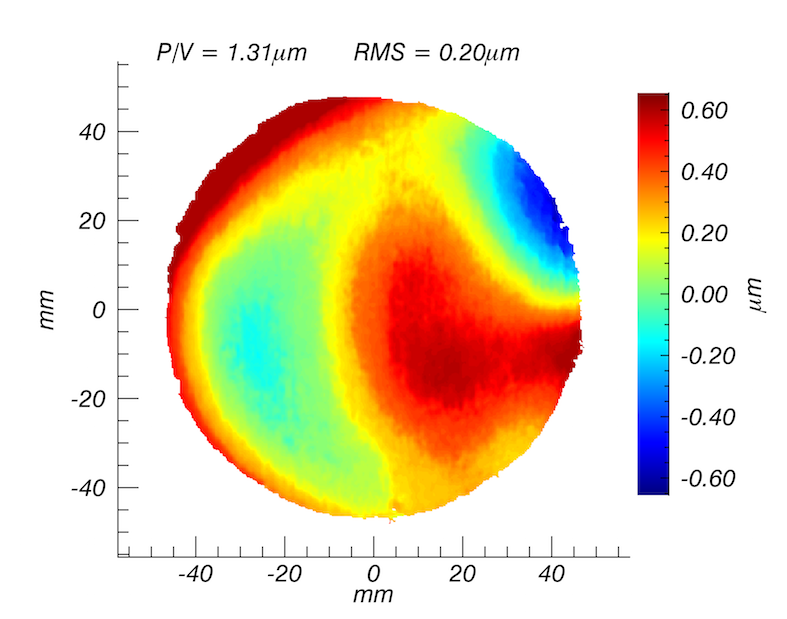}
}
\caption{{\it Left:} Picture of the collimator mirror test setup (all 3 together). The interferometer sits at the pupil position (the location of the diffraction grating). A collimated beam passes through the collimator in reverse (M3-M2-M1), goes through focus, and then is retro-reflected by a return sphere.  {\it Right:} Plots of the on-axis wavefronts of the full TMAs made up of ({\bf left}) the mirrors installed in SPIFFI from 2004-2016, and ({\bf right}) the new mirrors installed in the upgrade. The vertical striping in the wavefront of the old mirrors is from residual diamond turning marks on the mirror surfaces. The noticeable coma visible in the wavefront of the new mirrors is the result of an error in the surface of the new M3 mirror, which will be replaced during the upgrade to SPIFFIER. }
\label{fig:collimator}
\end{center}
\end{figure}

The spectrometer collimator is made up of three diamond-turned mirrors forming a classical Three Mirror Anastigmat (TMA): one sphere (M1), one convex off-axis prolate ellipse (M2), and one concave off-axis prolate ellipse (M3). Figure \ref{fig:collimator} shows the three mirrors set up in the lab on a mock cold-plate used for measuring the TMA wavefront. The mirror blanks are made of light-weighted thermally treated 5085 aluminum, which is the same material as the cold-plate, resulting in an athermal design. 

\begin{figure}[htbp!]
\begin{center}
\resizebox{1.0\textwidth}{!}{
\includegraphics[height=6cm,trim={0 0 0.5cm 0}, clip=true]{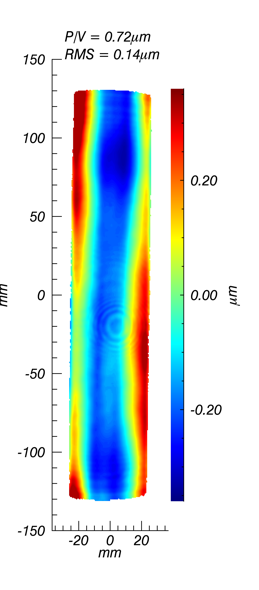}
\includegraphics[height=6cm,trim={0.5cm 0 0.4cm 0}, clip=true]{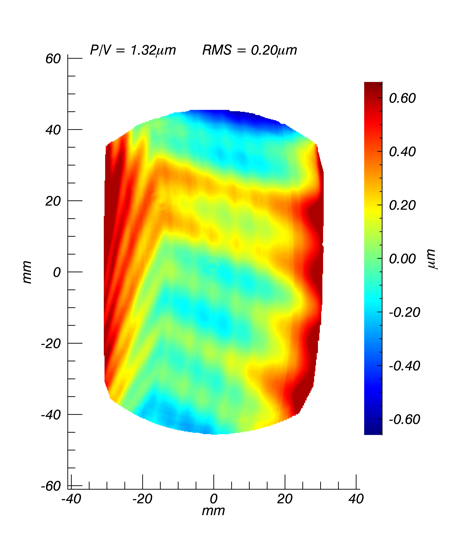}
\includegraphics[height=6cm,trim={0.5cm 0 0.4cm 0}, clip=true]{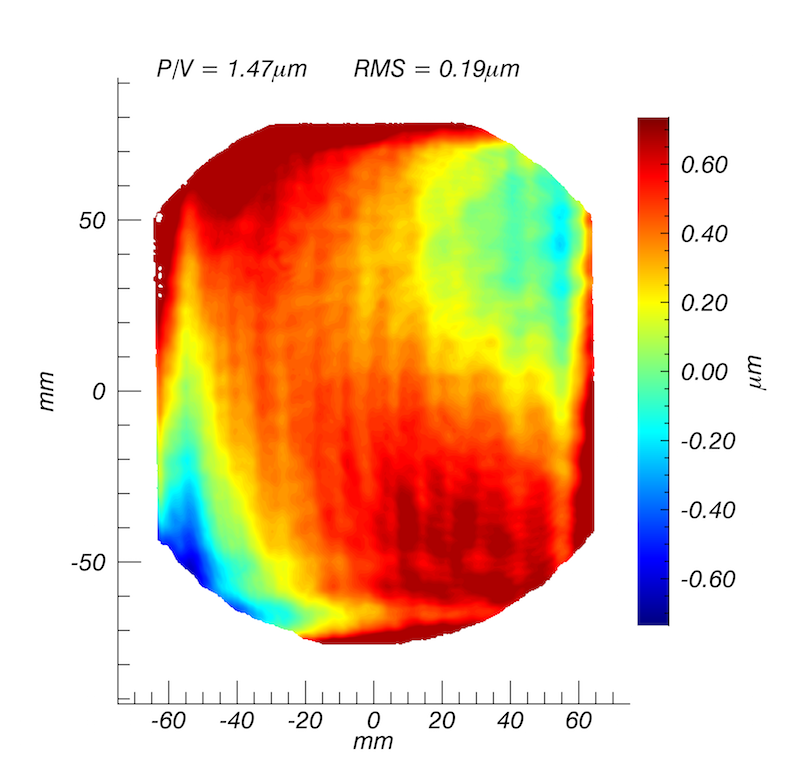}
}
\resizebox{1.0\textwidth}{!}{
\includegraphics[height=6cm,trim={0 0 0.5cm 0}, clip=true]{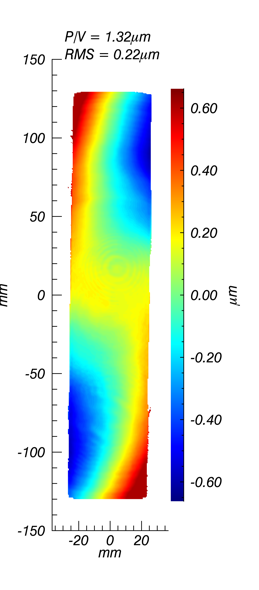}
\includegraphics[height=6cm,trim={0.5cm 0 0.4cm 0}, clip=true]{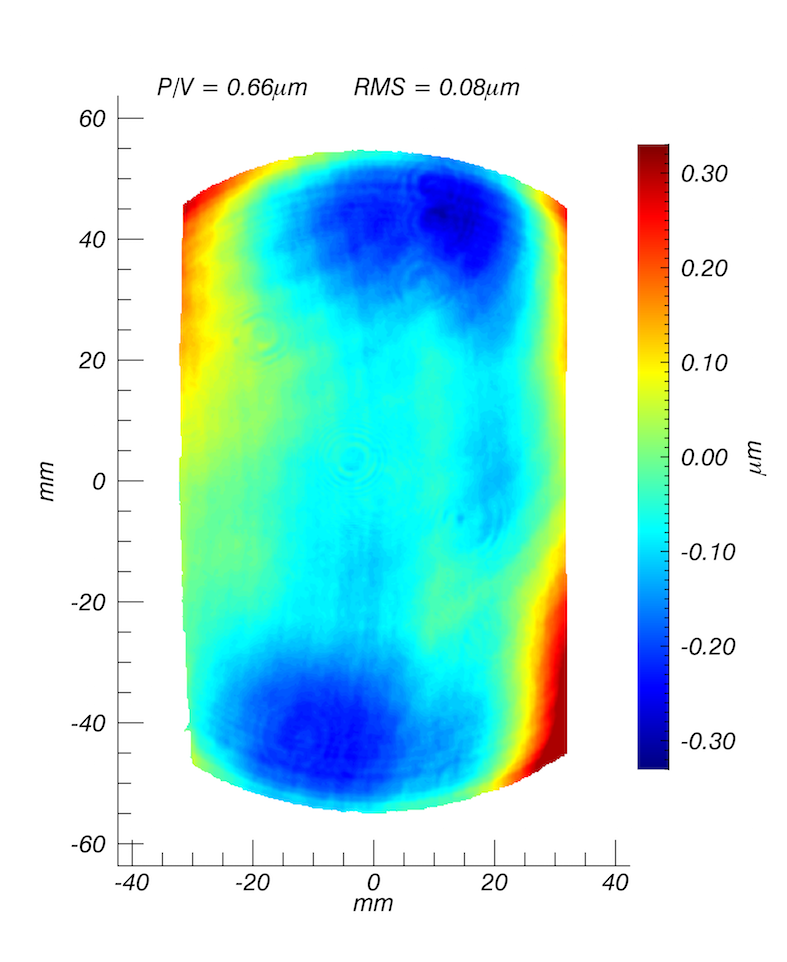}
\includegraphics[height=6cm,trim={0.5cm 0 0.4cm 0}, clip=true]{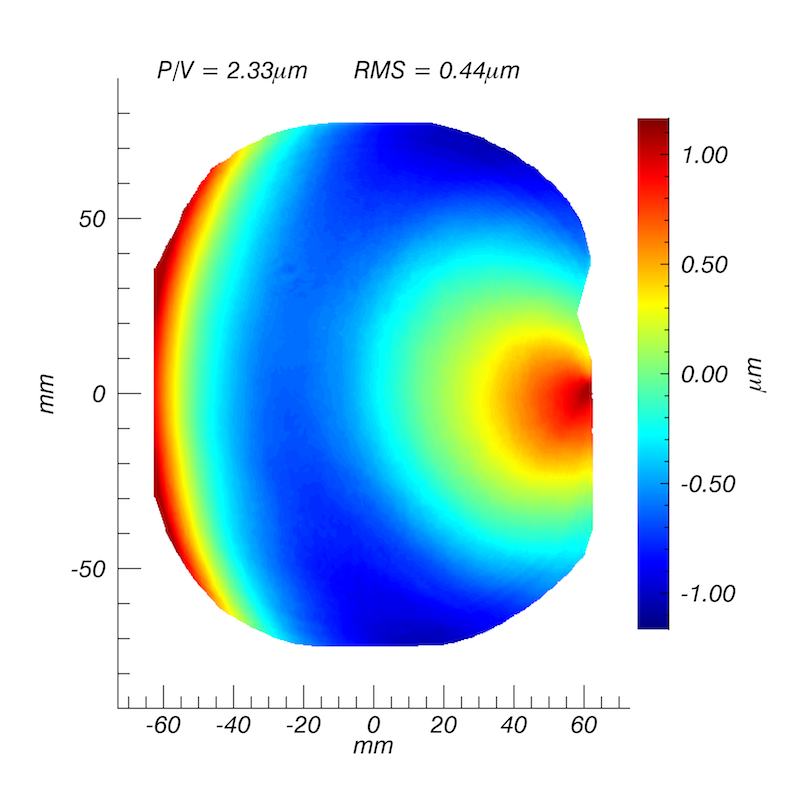}
}
\caption{Wavefronts of the individual collimator mirrors. {\it Left to right:} M1, M2, M3. {\it Top row:} The mirrors installed from 2005-2016 {\it Bottom row:} Mirrors installed in the 2016 upgrade. {\bf Note the different color scale on each wavefront plot.} As M1 is a spherical mirror, the post-polishing on the old mirrors was excellent--the wavefront is slightly better on the old mirror than the new one, though the footprint of any individual slitlet is very small on this mirror, so the large-scale variations do not matter as much. The old M2 mirror shows residual diamond turning marks (vertical stripes) as well as significant lap-polishing-like structure from the post-polishing, while the new mirror shows only very faint residual turning marks (PV \textless 10 nm) and much lower overall wavefront error. The old M3 mirror shows vertical striping and other residuals from machining and post-polishing as well as an over-all astigmatism, while the new mirror shows a wavefront error consistent with a wrong radius used in the machining process. The majority of the effect of this error on the new M3 can be removed through alignment of the M3 mirror. Note that due to the large 1D field, the footprints on the mirrors move significantly in the Y direction along the pseudo-slit. The result is that the overall TMA wavefront varies much more over the pseudo-slit with the old mirrors than the new mirrors (which are much smoother).}
\label{fig:wavefronts}
\end{center}
\end{figure}

The original mirrors in SPIFFI were manufactured in the years 2001-2003. They were coated with a layer of NiP, and then single-point diamond turned and post-polished. The mirrors were then coated in gold for higher reflectivity. Figure 6 of Eisenhauer et al 2003\cite{eisenhauer03} shows the significant residual turning marks on the collimator mirrors with a characteristic wavelength of 1 mm and depth of 200 nm PV. This resulted in diffraction at the turning marks, thus limiting the performance of the collimator and broadening the line-profiles of the instrument. Since the measurement shown in Eisenhauer et al 2003\cite{eisenhauer03}, a second set of mirrors was installed with more aggressive post-polishing in an attempt to reduce the amplitude of the turning marks. While the performance was better, the mirrors still had significant structure from both the turning and post-polishing which limited the performance of the collimator and resulted in a wavefront that varied over the pseudo-slit. This set of mirrors remained in SPIFFI from 2004-2016; unfortunately no measurements were available of the individual surfaces, so we were unable to directly compare these mirrors to our new ones before the upgrade. We measured the individual surfaces of these mirrors in the lab after the 2016 upgrade, and the wavefronts are shown in the top row of figure \ref{fig:wavefronts}. 

We manufactured new collimator mirrors for the 2016 upgrade. Given the rapid advances in machining control since 2001, the mirror manufacturer (Kugler) was able to provide surfaces within our specifications with low residual turning marks without the need for post-polishing. We decided to machine the mirror surfaces on the bare 6061 aluminum blanks with no NiP coating to eliminate the risk of potential bimetallic bending stresses when the mirrors were cooled to 80K. The mirrors were then coated with a thin layer of chrome, a thicker layer of gold, and finally a protection layer. The resulting surface roughness was $\sim$5 nm (acceptable for the J-K band wavelength range), with nearly invisible residual turning marks. The bottom row of figure \ref{fig:wavefronts} shows the overall wavefronts of each of the new mirrors.  The protection layer additionally decreases the throughput slightly at short wavelengths. The reflectivity of the new surfaces ranges from 96.5\% to $\gt$98.5\% over the J-K bands, compared to a reflectivity of $\gt$98.5\% for the bare-gold old mirrors. 

The right panel of figure \ref{fig:collimator} shows the overall on-axis TMA wavefront for the old and new mirrors. The dominant error in the wavefront of the new TMA is a coma term, which is the result of a parameter error in the manufacturing of the surface of M3. With the nominal mirror alignment, this error resulted in $\sim$6 waves of 45 degree astigmatism in the TMA. This was removed by changing the alignment of M3, however, $\sim$ 350nm of coma remains. A new M3 mirror is in manufacturing, and will be installed for SPIFFIER in ERIS. Despite this, the performance of the new mirrors is superior to the old ones, resulting in cleaner, less variable line profiles across the pseudo-slit. Section \ref{sec:resolution} shows an analysis of the line profiles of the instrument before and after the upgrade.  More information on the individual mirror measurements and alignment is available in Gr\"aff 2016\cite{graeff16}.

%%%%%%%%%%%%%%%%%%%%%%%%%%%%%%%%%%%%%%%%%%%%%%%%%%%%%%%%%%%%%
\subsection{Other changes}
\label{sec:otherchanges}

During the course of the early upgrade, we exchanged the field stop baffle in the sky spider housing for a larger one. The original baffle was designed to exactly bracket the large FOV, and imperfect positioning of the pre-optics wheel occasionally resulted in additional vignetting of slitlet numbers 1, 2, 31, or 32. The larger baffle prevents this from occurring.

When opening the cryostat, we found that the grating wheel drive gear assembly as well as the large toothed wheel on the grating wheel were damaged and worn from use. We replaced the toothed wheel and grating drive gear assembly with spares, and are currently considering a re-design of the grating drive assembly for SPIFFIER in ERIS.

%%%%%%%%%%%%%%%%%%%%%%%%%%%%%%%%%%%%%%%%%%%%%%%%%%%%%%%%%%%%%
\section{Performance after the early upgrade}
\label{sec:performance}

The performance of the instrument was improved during this upgrade. The Strehl Ratio achievable during operations was improved by a combinations of three things 1) The new pre-optics have slightly better image quality than the old ones, 2) During the SPIFFI upgrade, maintenance was performed on MACAO that improved its performance\cite{cortes16}, and 3) The new collimator mirrors have lower wavefront variation over the FOV, reducing PSF distortions. The spectral line profiles were improved by the replacement of the collimator mirrors, as the wavefront deformations in the old mirrors, particularly M2, spread the line profiles and produced variation in the line profiles over the detector. Finally, due to modern filter and AR coating technology, the throughput of the instrument was improved by the replacement of the filters and pre-optics. The following sub-sections report on the performance gains after the upgrade.

%%%%%%%%%%%%%%%%%%%%%%%%%%%%%%%%%%%%%%%%%%%%%%%%%%%%%%%%%%%%%
\subsection{Image quality}
\label{sec:image}

The Strehl Ratio achievable depends mainly on the quality of the pre-optics and the quality of the AO correction. The new pre-optics installed during the upgrade have near optimal performance, with SRs measured in laboratory testing near 100\% (see section \ref{sec:preoptics}). Additionally, the MACAO maintenance improved the on-sky performance of the AO\cite{cortes16}. Due to the size of the image slicer slitlets, the measurement of point source PSFs within SPIFFI is only well sampled in the 25mas pixel scale. Table \ref{tab:strehl} lists the measured Strehl ratios and PSF FWHMs in the 25 mas pixel scale using the calibration fiber. Using the fiber eliminates the effects of atmospheric conditions in the measurement of performance, and thus acts as an upper-limit on performance. The measured SRs on the calibration fiber are slightly improved (order 5-10\%) over before the upgrade.

\begin{table}[htb]
\begin{center}
\begin{tabular}{ccccc}
& J  & H & K & H+K \\
\hline
\hline
&\includegraphics[width=3.5cm]{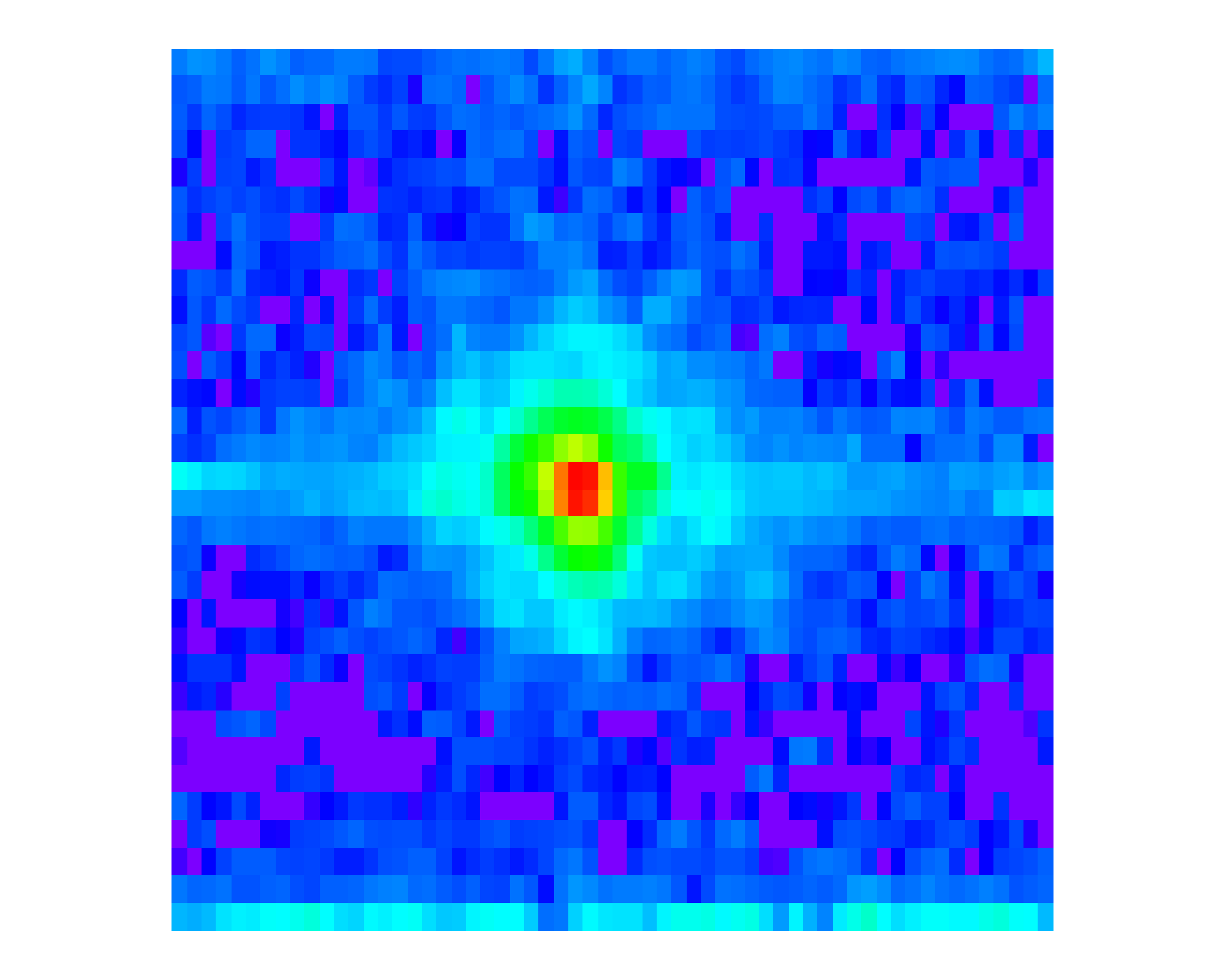}  &\includegraphics[width=3.5cm]{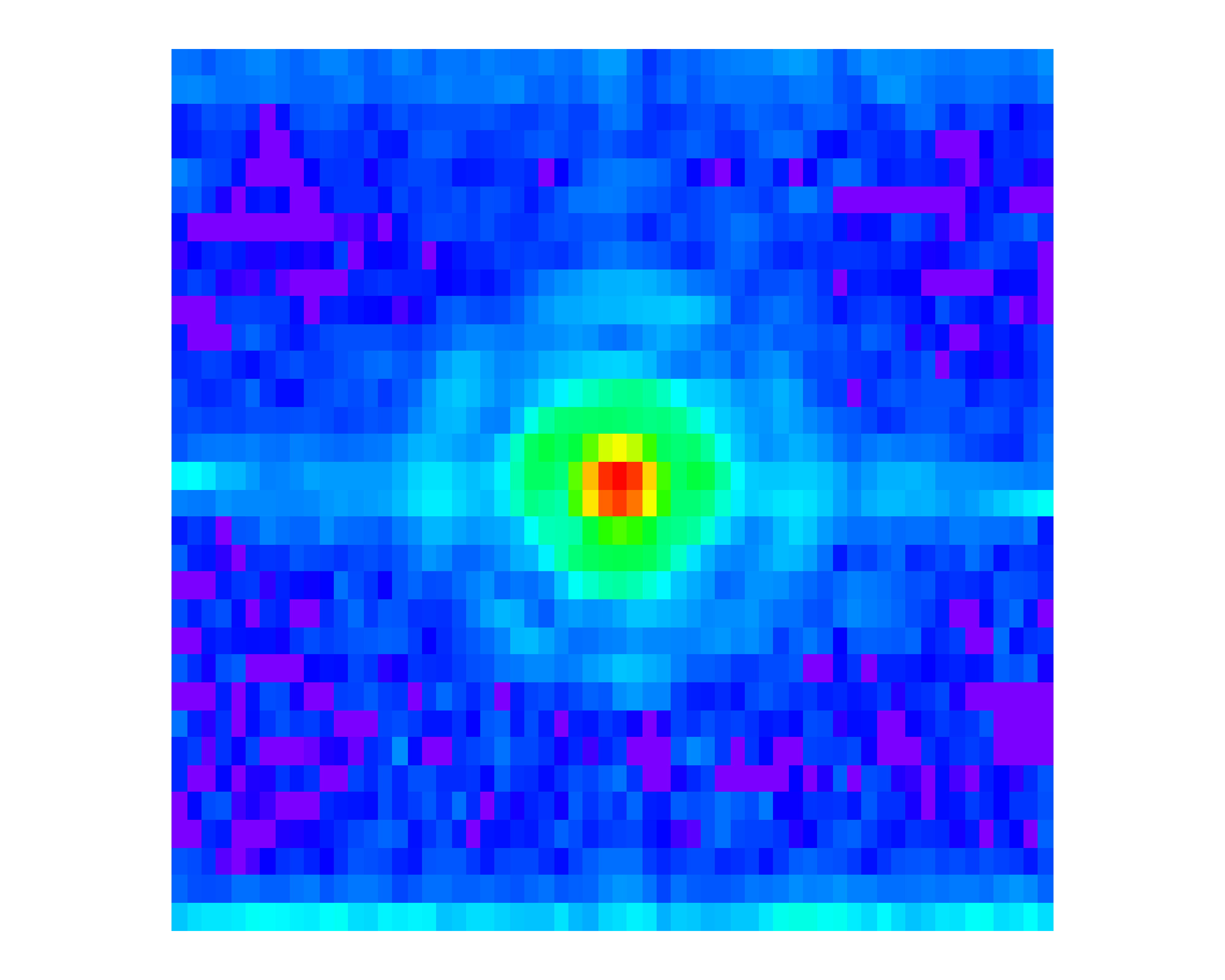}  & \includegraphics[width=3.5cm]{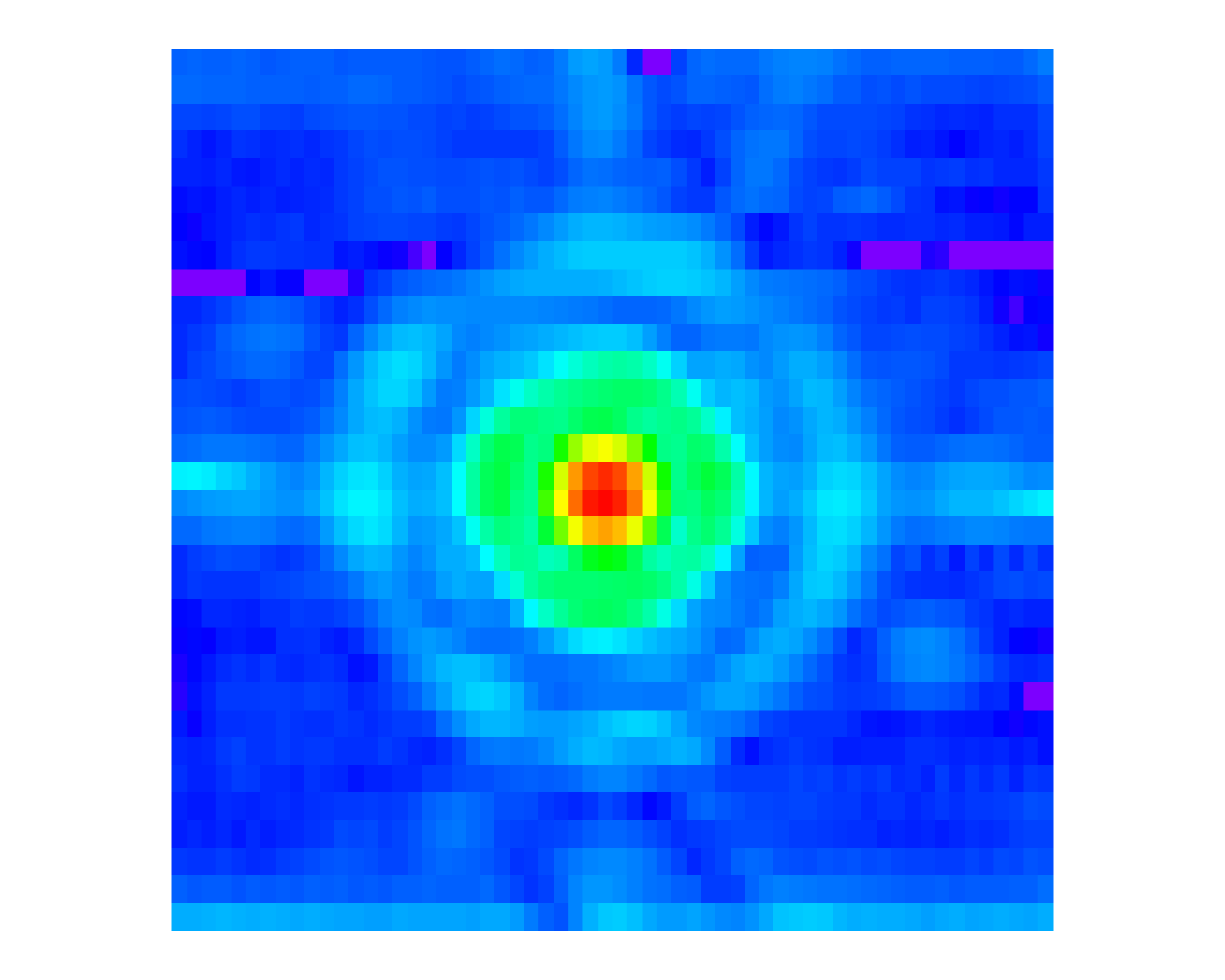} &\includegraphics[width=3.5cm]{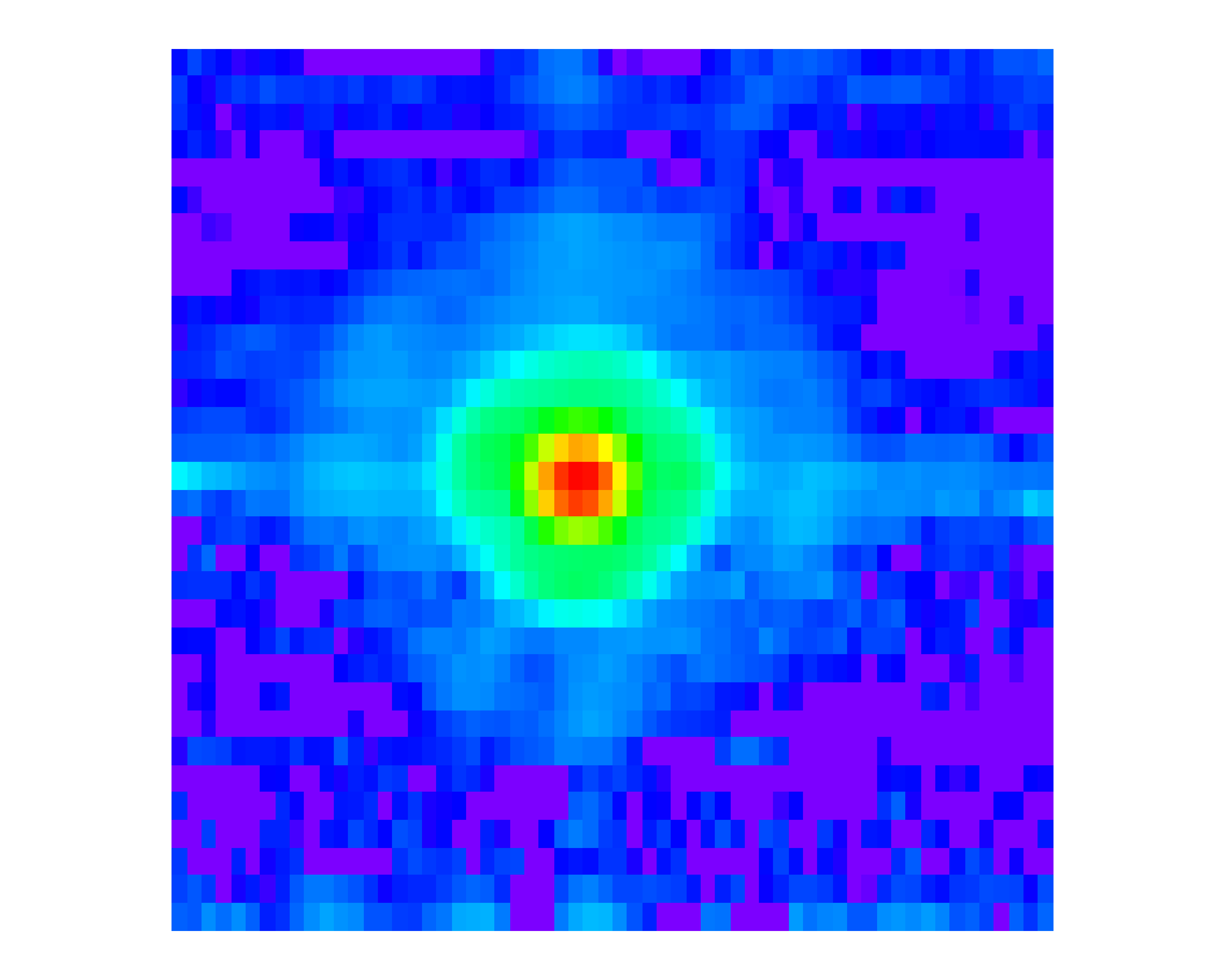}  \\
SR &58 $\pm$ 17 \%&83 $\pm$10 \%&95 $\pm$ 6 \%&78 $\pm$ 7 \%\\
FWHM & 44 mas&46 mas&56 mas&57 mas\\
 \hline
\end{tabular}
\vspace{0.05in}
\caption{Summary of the Strehl Ratios (SR) and PSF size in the 25 mas pixel scale as measured using the calibration fiber to decouple performance from atmospheric conditions. The error bar comes from the fact that the PSF is nearly Nyquist sampled.\cite{roberts04} Additionally, we have achieved on-sky SRs of (34, 44, 69, 63) \% in (J, H, K, H+K) bands respectively).}
\label{tab:strehl}
\end{center}
\end{table}

The on-sky on-axis NGS Strehl Ratios achieved vary from night to night, but in the best conditions approach the values listed in table \ref{tab:strehl}. Since the upgrade, we have achieved on-sky SRs of (34, 44, 69, 63) \% in (J, H, K, H+K) bands respectively. Additionally, the performance when using the Laser Guide Star (LGS) with an off-axis NGS was improved by the replacement of the MACAO membrane mirror during the upgrade\cite{cortes16}.
%%%%%%%%%%%%%%%%%%%%%%%%%%%%%%%%%%%%%%%%%%%%%%%%%%%%%%%%%%%%%
\subsection{Line Profiles and Spectral Resolution}
\label{sec:resolution}

The line profiles in SPIFFI are asymmetric with shoulders that deviate from a Gaussian profile, and shapes that vary over the detector. Thatte et al. 2012\cite{thatte12} describes a method for measuring super-sampled line profiles using OH lines. We use this method to measure the line profiles of the instrument before and after the upgrade using the spectral lines of the arc lamps in the calibration unit. We report the resolution of the instrument and improvement after the upgrade in section \ref{sec:specres}, and the measured super-sampled profiles in section \ref{sec:measline}. We used additional measurements and optical modelling to finally determine the root cause of the SPIFFI line profile shapes and propose a fix for the upgrade to SPIFFIER in ERIS in section \ref{sec:lineexplain}.

%%%%%%%%%%%%%%%%%%%%%%%%%%%%%%%%%%%%%%%%%%%%%%%%%%%%%%%%%%%%%
\subsubsection{Spectral resolution}
\label{sec:specres}

\begin{figure}[htbp!]
\resizebox{1.0\textwidth}{!}{
\begin{floatrow}
\capbtabbox{%
\begin{tabular}{cccccc}
\hline
{\small Band}&{\small Scale}&{\small FWHM}&{\small Resolution}&{\small Improvement}\\
 & [mas]& [pixels] & $\lambda_c/d\lambda$& [\%]\\
\hline
\hline
J & 25&3.7 &2270 & -6 \\
 & 100&4.4 &1880 & 7\\
 & 250&4.3 &1960 & 5\\
 \hline
H & 25&2.7 &3090  & -10\\  
 & 100&3.2 &2570  & 11\\
 & 250&3.0 &2710  & 11\\
 \hline
K & 25&1.6 &5330  & 8\\ 
 & 100&1.9 &4670  & 34\\
 & 250&2.2 &4070  & 17\\
 \hline
H+K&25&1.9 &2120  & -6\\
 & 100&2.0 &1940 & 33\\
 & 250&2.3 &1730 & 22\\
 \hline
\end{tabular}
}{%
\caption{Summary of the instrument resolution after the upgrade with standard detector sampling. The design FWHM of the lines in detector pixels is $\sim$1.6. The trend is for larger FWHM at shorter wavelengths and larger pixel scales. The FWHM and resolution varies across the detector, so the number reported is the average for the band. The error on this measurement is $\sim$5\%.}
\label{tab:specres}
}
\ffigbox{%
\includegraphics[width=0.5\textwidth]{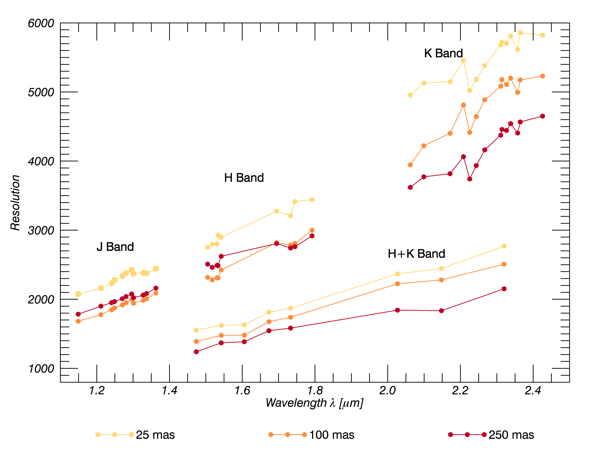}
}{%
  \caption{Plot of the spectral resolution as a function of wavelength. The three pixel scales (25, 100, and 250 mas) are plotted in (yellow, orange, and red) for each grating. The resolution decreases in larger pixel scales and shorter wavelengths.}%
  \label{fig:specres}
}
\end{floatrow}
}
\end{figure}

The spectral resolution of the instrument was measured in each band and pixel scale before and after the upgrade. The resolution is listed in table \ref{tab:specres}, and was calculated as $\lambda_c/d\lambda$, where $\lambda_c$ is the center wavelength of the band and $d\lambda$ is the FWHM of the Gaussian fit to the arc-lamp spectral lines using the standard detector sampling. 
The gains in resolution of $\sim$10-30\% are mostly in the 100 and 250 mas pixel scales, where the beam footprints on the collimator mirrors are larger, and thus are affected more strongly by the residual diamond turning marks. We attribute this increase in resolution in the large pixel scales after the upgrade to the change of the collimator mirrors. Figure \ref{fig:specres} shows the resolution ($\lambda/d\lambda$) as a function of wavelength for each pixel scale and diffraction grating. 

%%%%%%%%%%%%%%%%%%%%%%%%%%%%%%%%%%%%%%%%%%%%%%%%%%%%%%%%%%%%%
\subsubsection{Measured Super-sampled line profiles}
\label{sec:measline}

\begin{figure}[htbp!]
\begin{center}
\resizebox{1.0\textwidth}{!}{
\includegraphics[width=1.0\textwidth]{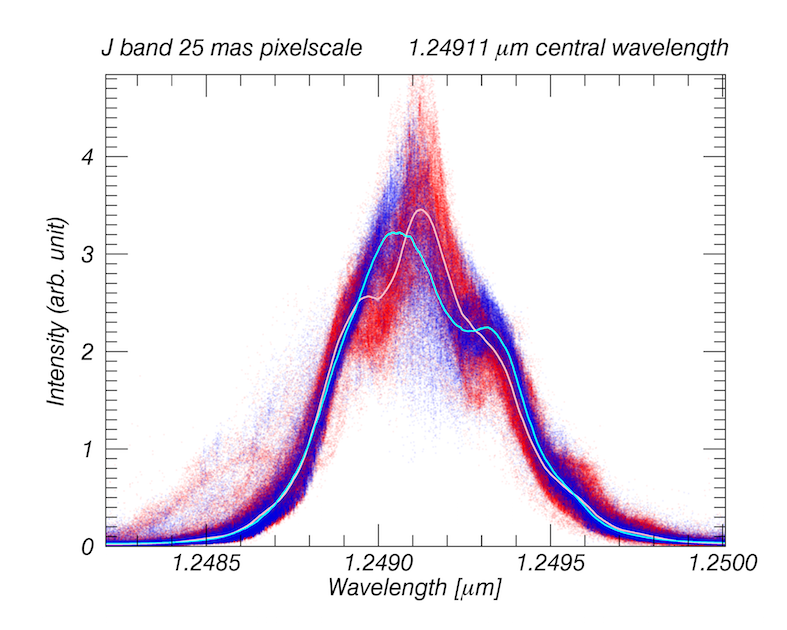}
\includegraphics[width=1.0\textwidth]{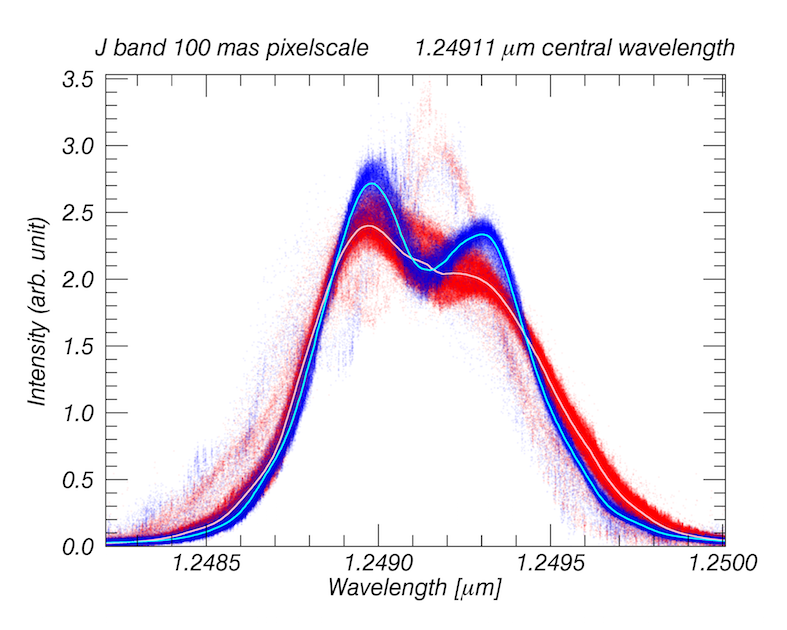}
\includegraphics[width=1.0\textwidth]{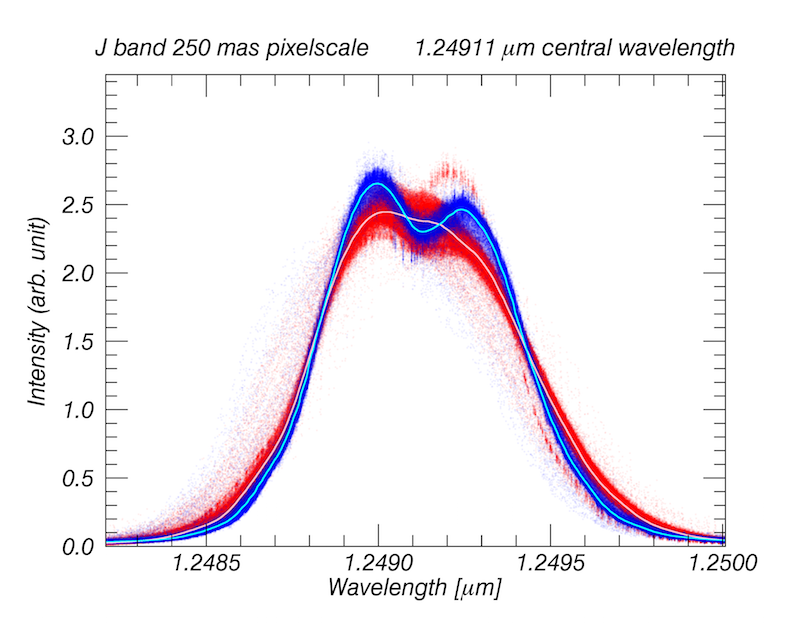}
}
\resizebox{1.0\textwidth}{!}{
\includegraphics[width=1.0\textwidth]{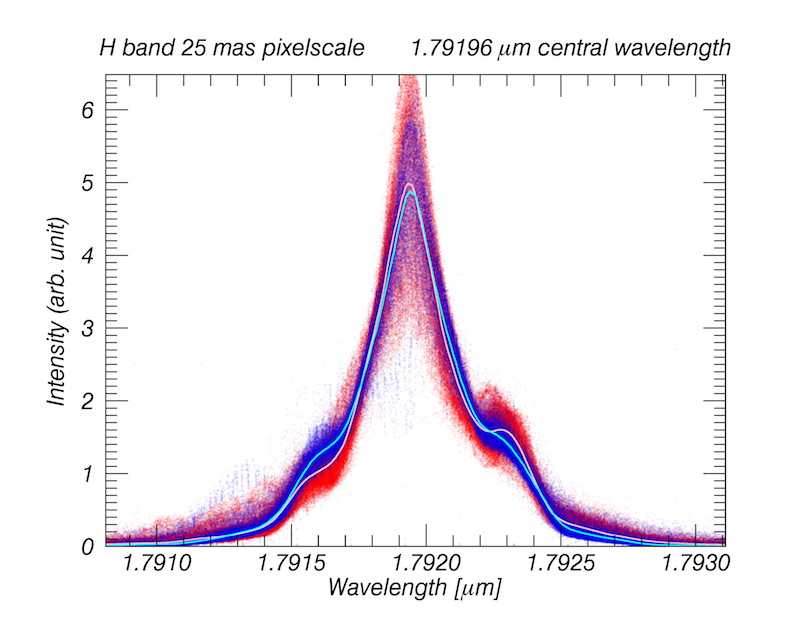}
\includegraphics[width=1.0\textwidth]{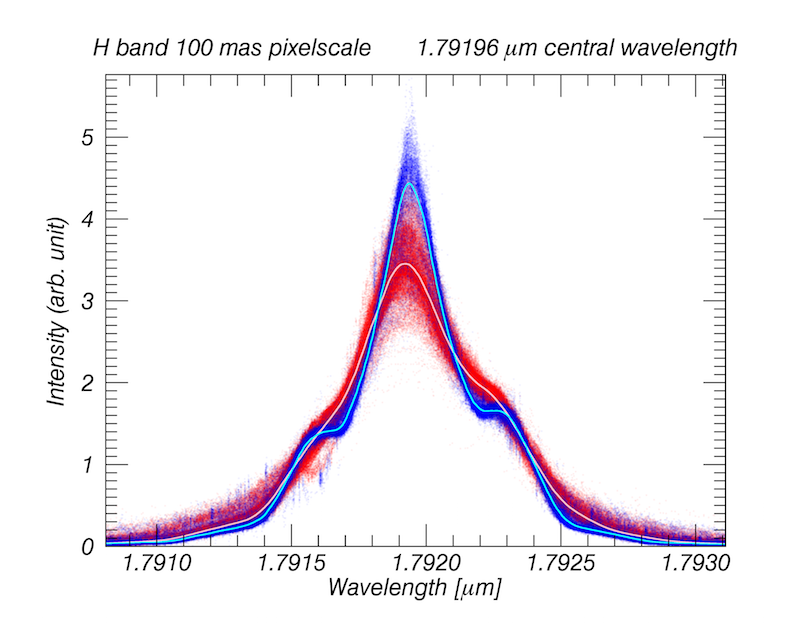}
\includegraphics[width=1.0\textwidth]{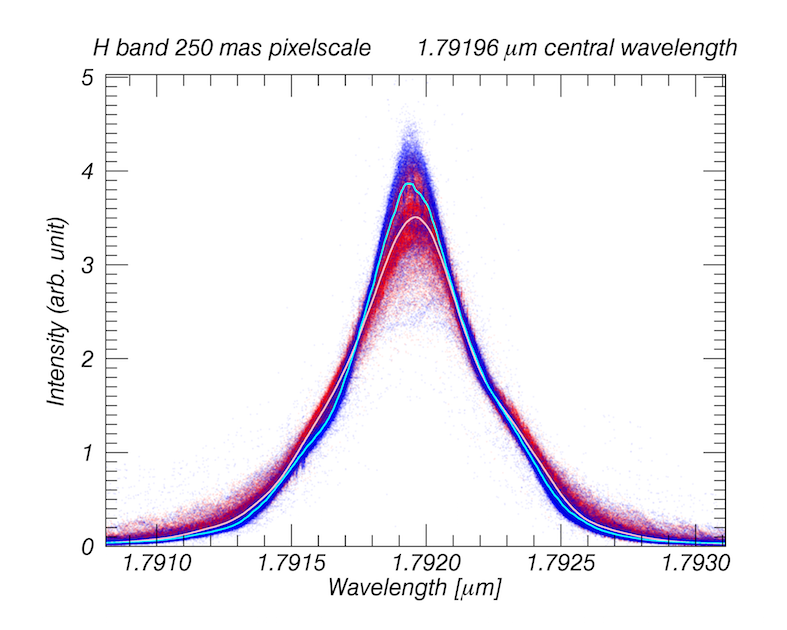}
}
\resizebox{1.0\textwidth}{!}{
\includegraphics[width=1.0\textwidth]{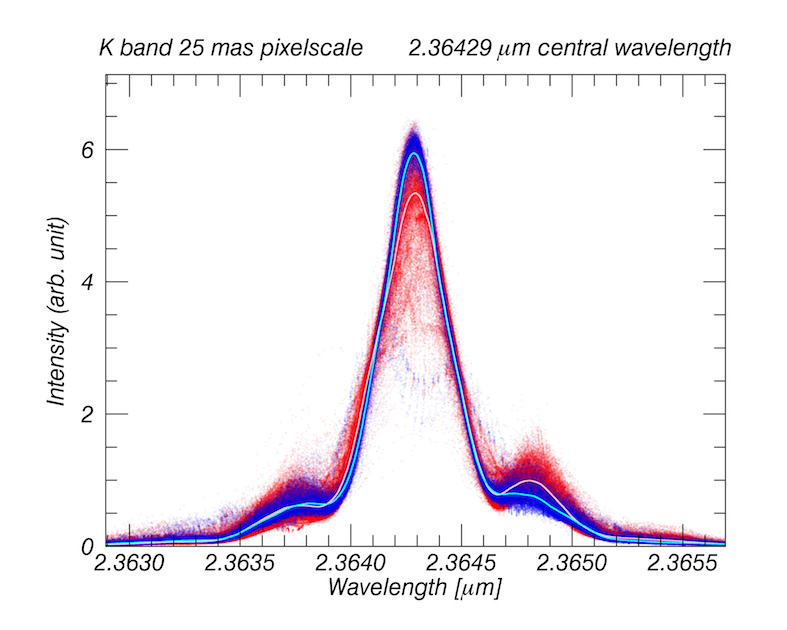}
\includegraphics[width=1.0\textwidth]{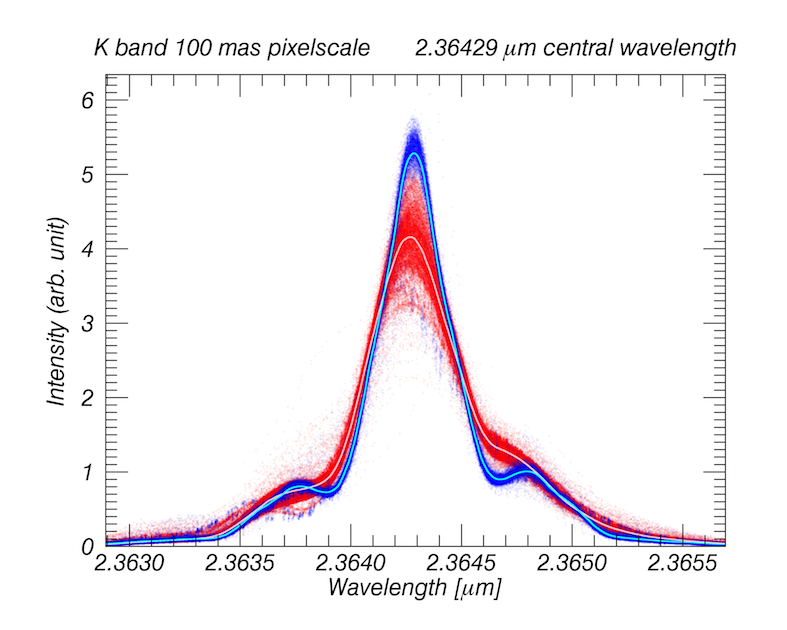}
\includegraphics[width=1.0\textwidth]{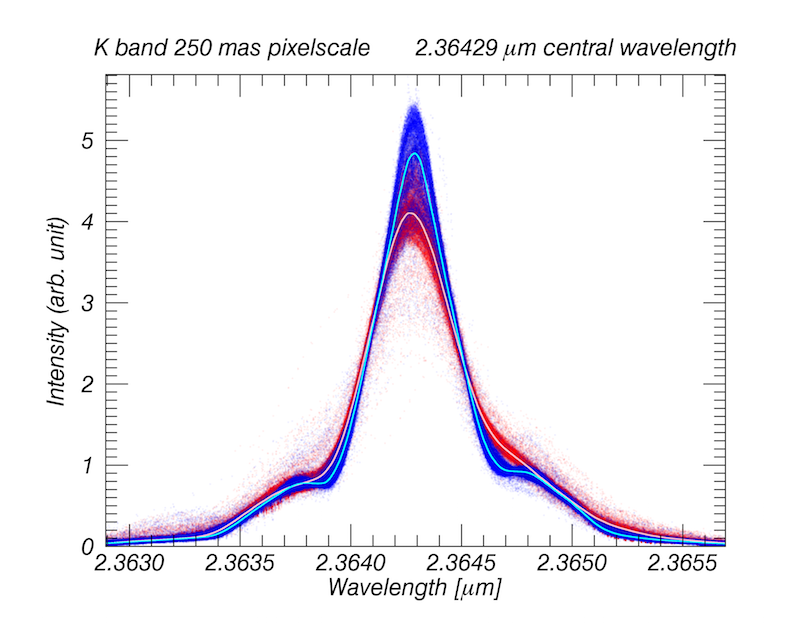}
}
\resizebox{1.0\textwidth}{!}{
\includegraphics[width=1.0\textwidth]{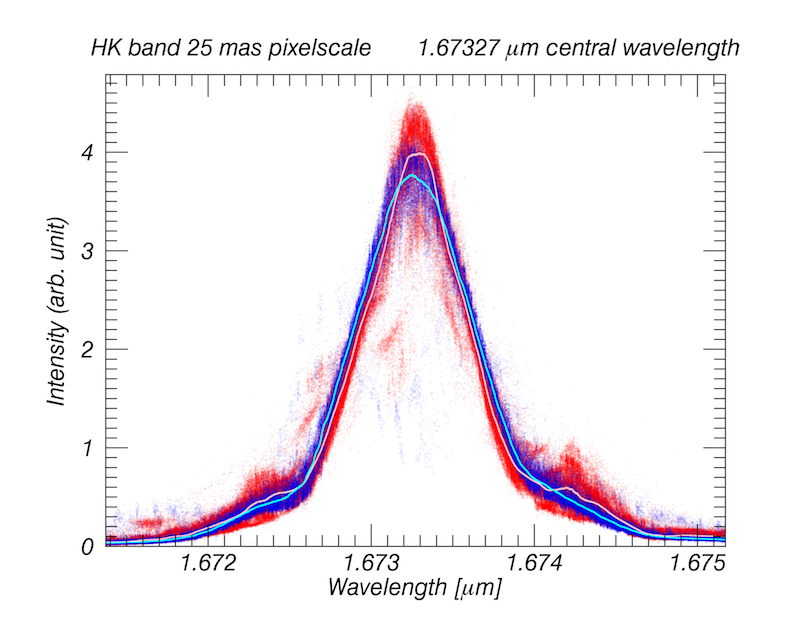}
\includegraphics[width=1.0\textwidth]{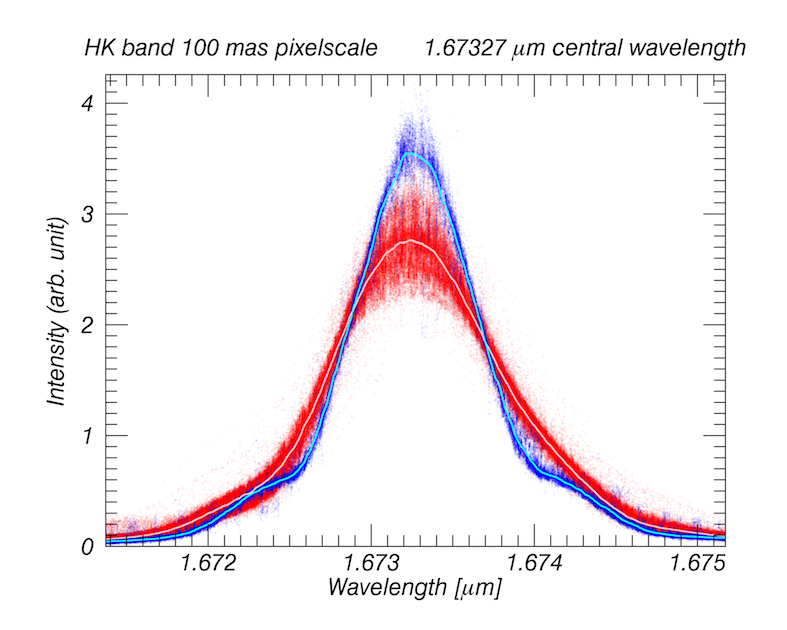}
\includegraphics[width=1.0\textwidth]{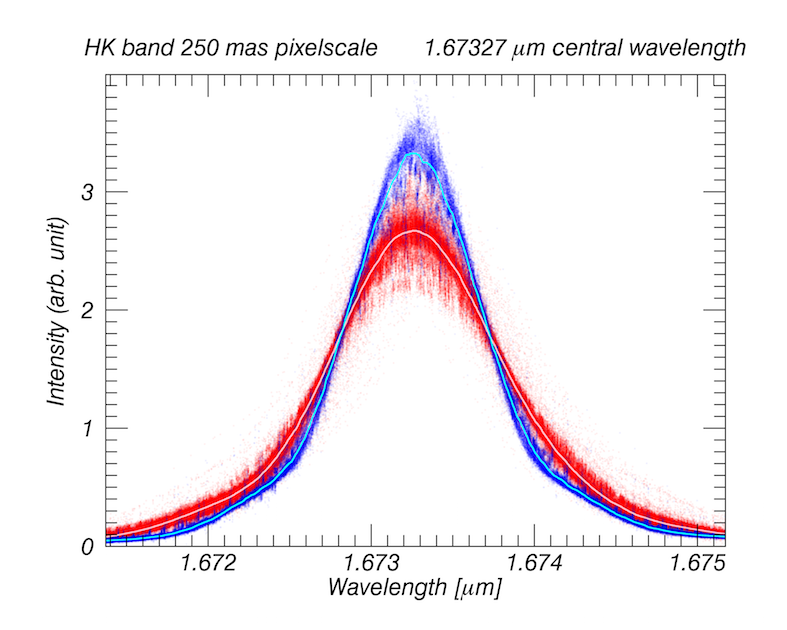}
}
\caption{Average line profile at a single wavelength in each band and pixel scale. {\it Rows, Top to bottom}: J, H, K, H+K gratings. {\it Columns, Left to Right}: 25, 100, 250 mas pixels scales. The cloud of points represents all data points of the supersampled lines from all locations on the detector, while the solid lines are the average of the clouds. {\it Blue}: Post-upgrade {\it Red}: Pre-upgrade.}
\label{fig:lineprofiles}
\end{center}
\end{figure}

The super-sampled line profiles have a distinct shape and vary with wavelength and pixel scale. The trend in wavelength is for the line profiles to have small shoulders in the long wavelengths of K-band, which when moving to shorter wavelengths gradually rise up, until in J-band, the shoulders are nearly as high as the central peak. In the 25 mas pixel scale, the shoulders are easily distinguishable (left column of figure \ref{fig:lineprofiles}, while in the larger 100 mas and 250 mas pixel scales the shoulders are washed out, and the line profile only appears broadened, without distinct peaks (middle and right columns of figure \ref{fig:lineprofiles}). Figure \ref{fig:lineprofiles} shows the average line profile of the instrument before (red) and after (blue) the upgrade in all pixel scales at a single wavelength near the center of each band.

The line profiles additionally vary across the detector in two ways. There is a variation across a single slitlet, which we attribute to the fact that the slitlet mirrors are each tilted on the small slicer, and thus the edges of the slitlets are out of focus with respect to the center of the slitlets. In general we see higher resolutions in the centers of the slitlets. Each slitlet in the small slicer has a different angle, and thus this variation is different in each slitlet. The upgrade did not significantly change the line profile variation across a single slitet. There is also a variation in line profile across the full length of the pseudo-slit, which comes from each slitlet having a different beam footprint on the collimator mirrors and diffraction grating. The footprints in different slitlets are almost completely separate on M1, overlap slightly on M2, overlap significantly on M3, and are almost completely overlapping on the grating. The replacement of the collimator mirrors reduced the variation in the average line profile in each slitlet over the full length of the pseudo-slit, which we attribute to the much greater uniformity of the new M2 mirror. Plots of the variability of the line profiles over single slitlets and over the pseudo-slit are available in Gr\"aff 2016.

%%%%%%%%%%%%%%%%%%%%%%%%%%%%%%%%%%%%%%%%%%%%%%%%%%%%%%%%%%%%%
\subsubsection{Explanation for line shapes}
\label{sec:lineexplain}

The line profiles in the SPIFFI instrument have been the subject of investigation for many years. Iserlohe 2004\cite{iserlohe04} details the investigations into the line profiles of the instrument at the time of integration. The different line widths in the H and K bands was thought to be  uncompensated axial color in the spectrometer camera. The smearing on the line profiles in all wavelengths was assumed to be the collimator mirrors, but after they were replaced by the mirrors installed from 2004-2016, there was only minimal improvement, though those mirrors were also not ideal (see measurements in section \ref{sec:mirrors}). Finally after measurements with the new 2k camera, the J-band line profiles were blamed on ``bad surface quality" of the J-band grating. 

After the replacement of the collimator mirrors in this upgrade, we were able to measure the wavefronts of the old collimator mirrors. It became clear from both these measurements, and the only small observed increase in spectral resolution after the upgrade, that the collimator mirrors were only a minor contributing factor to the line profiles in the instrument. Additionally, the resolution variation with wavelength was consistent between the old 1K camera and the current 2K camera that was installed in 2004. The only remaining factor that could be affecting the line profiles is the diffraction gratings.  

\begin{figure}[htbp!]
\begin{center}
\resizebox{1.0\textwidth}{!}{
\includegraphics[width=1.0\textwidth]{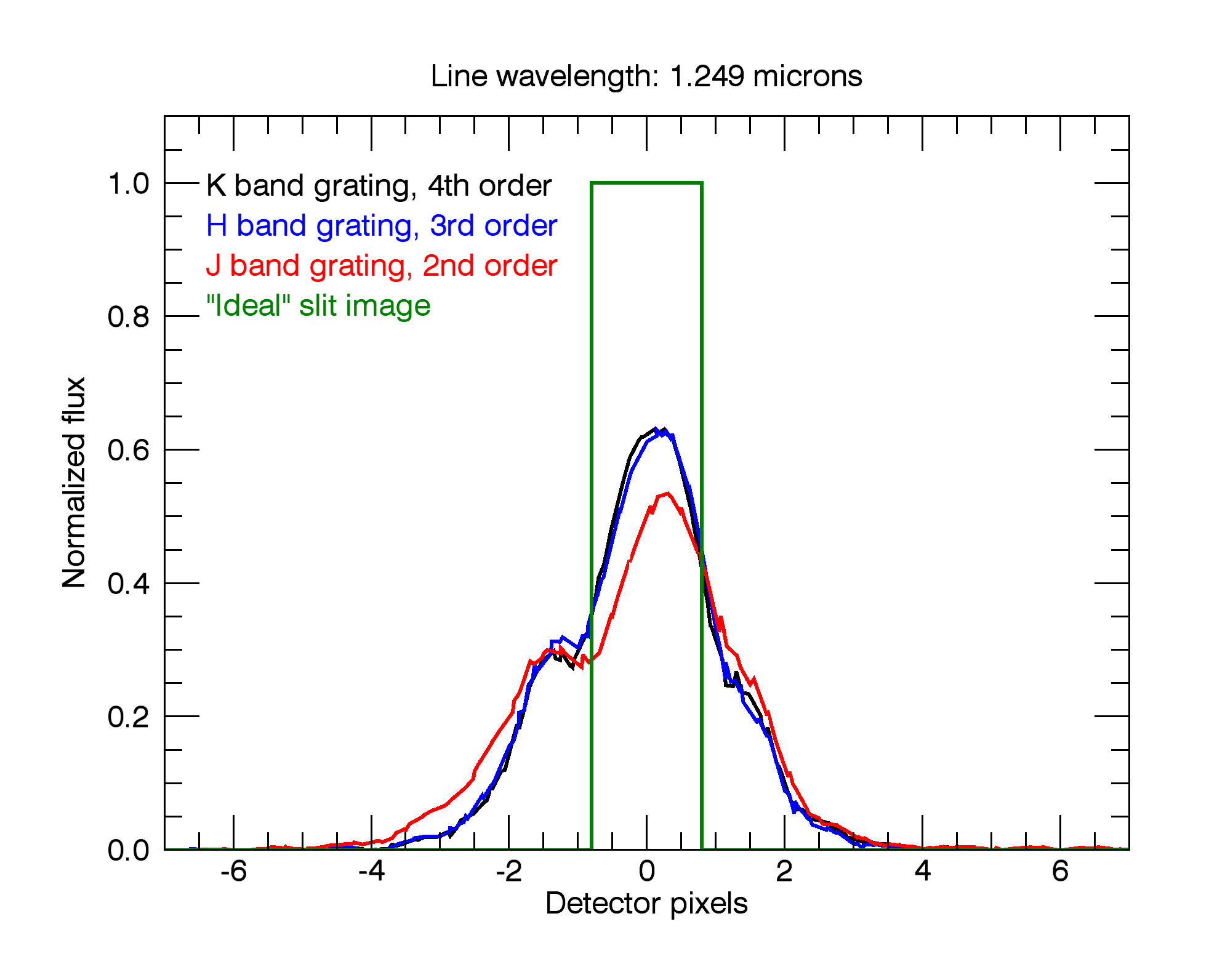}
\includegraphics[width=1.0\textwidth]{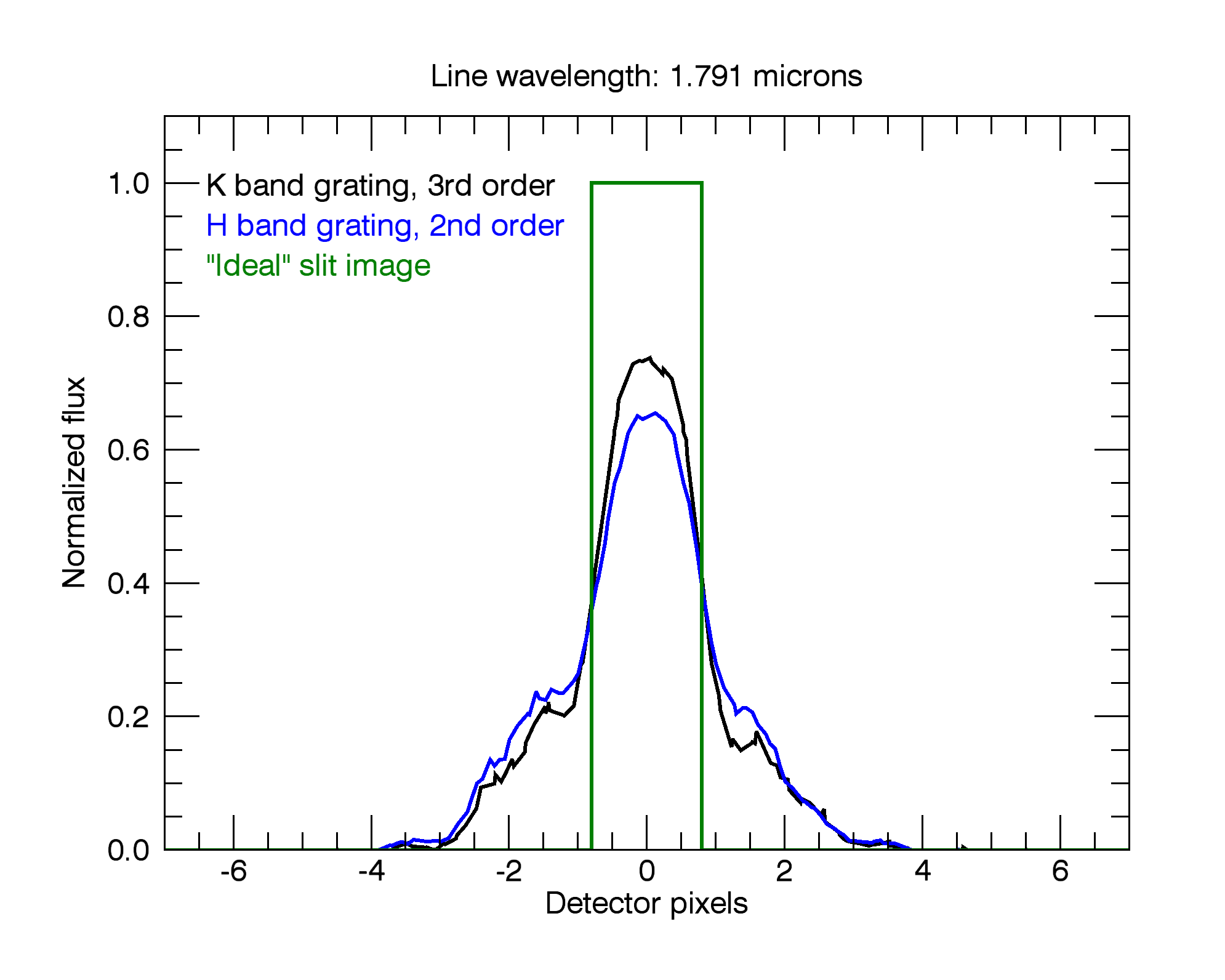}
\includegraphics[width=1.0\textwidth]{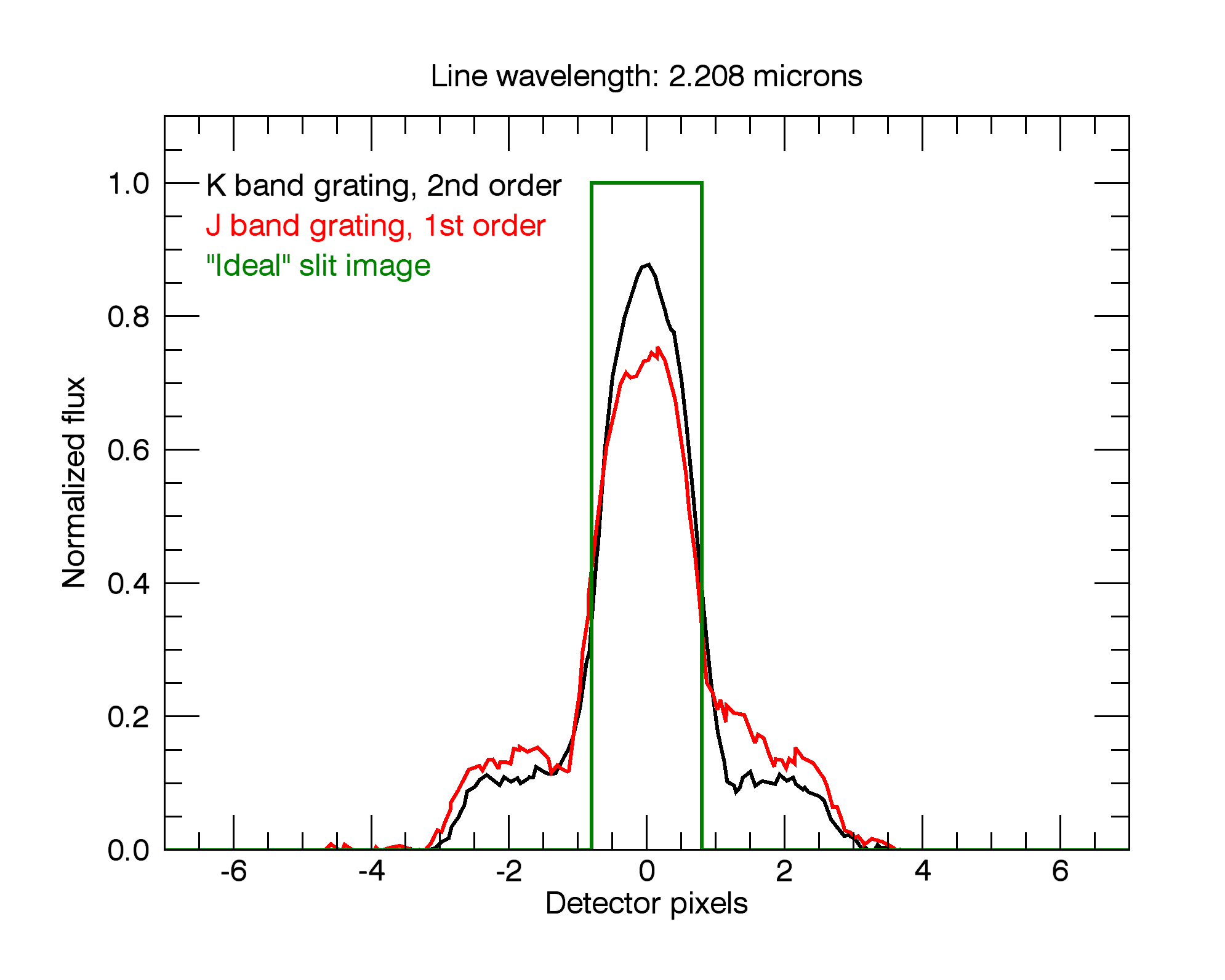}
}
\caption{ Plots of single super-sampled lines on different diffraction gratings in the 25 mas pixel scale, normalized to the total integrated power in the line. {\it Left}: $\sim$1.25 micron line using the J-, H-, and K-band diffraction gratings. {\it Middle}: $\sim$1.8 micron line using the H- and K-band gratings. {\it Right}: $\sim$2.2 micron line using the J- and K-band diffraction gratings. Note that the data is plotted as recorded on the detector (i.e., not wavelength calibrated). The line morphology is remarkably similar between the gratings, as expected by our optical model, though not identical.}
\label{fig:differentorders}
\end{center}
\end{figure}

We performed an experiment in SPIFFI in which we obtained super-sampled line profiles in different diffraction orders of the gratings to obtain lines with wavelengths in the J-K bands on each diffraction grating. For example, using the K-band grating in 4th order, the H-band grating in 3rd order, and the J-band grating in 2nd order (the default), we were able to measured J-band lines on each grating. Figure \ref{fig:differentorders} shows $\sim$1.25, $\sim$1.8, and $\sim$2.2 micron lines measured on the J-, H- and K-band gratings. We chose to use the 25mas pixel scale, since the effects of large-scale wavefront errors would affect the resulting line profiles less. The three diffraction gratings give remarkably similar results, and clearly show the progression of the shoulder height with wavelength is as expected from the optical model. From this we infer that if the diffraction gratings are causing the line profile variability, each diffraction grating must have a wavefront error which affects the line profiles in the same way.

\begin{figure}[htbp!]
\begin{center}
\resizebox{0.8\textwidth}{!}{
\includegraphics[height = 6cm]{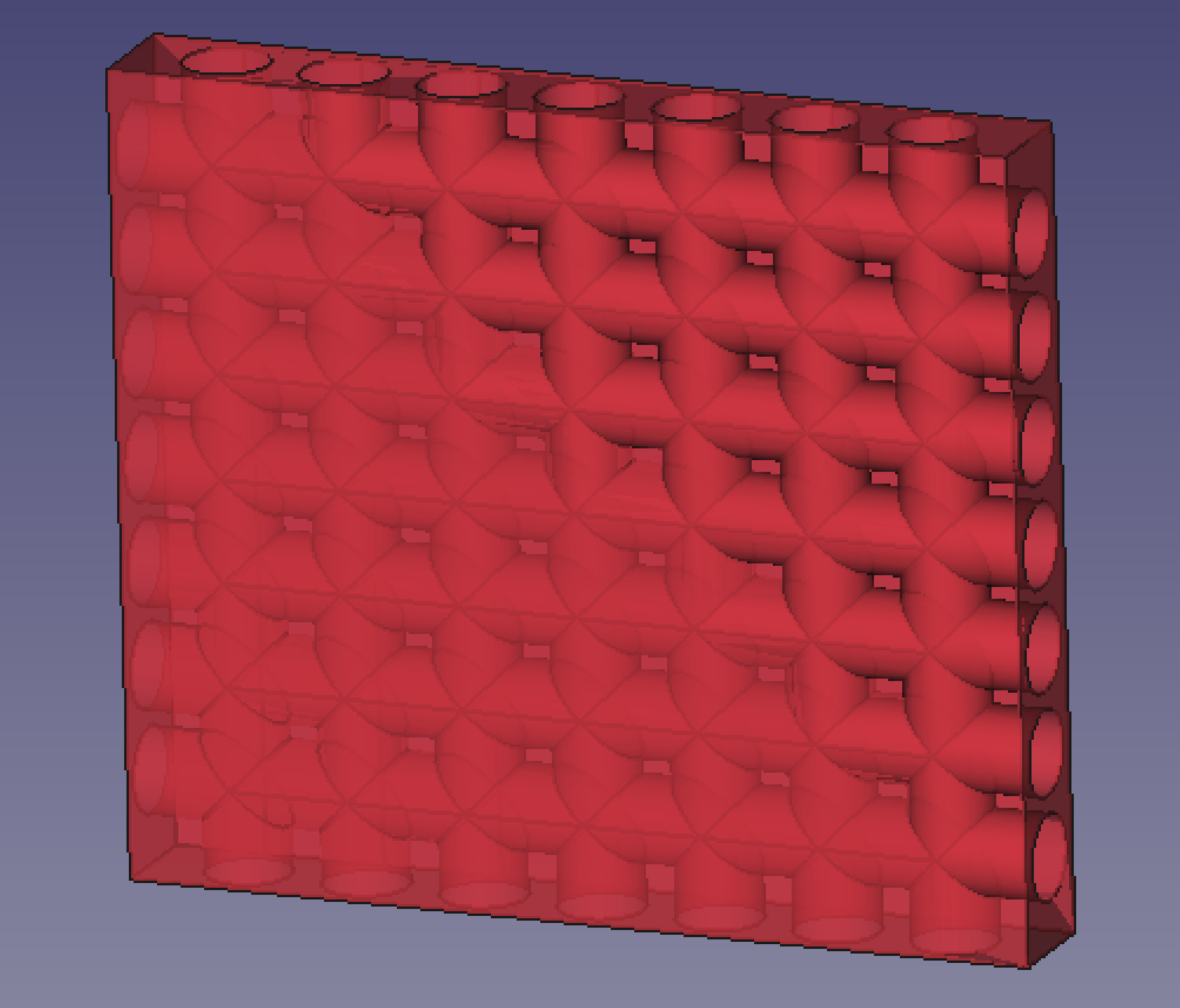}
\includegraphics[height = 6.5cm]{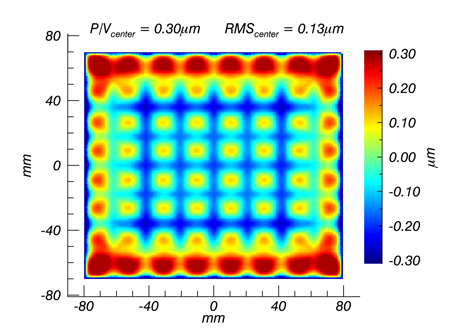}
}
\caption{Left: CAD model of grating blank with lightweighting holes. Right: Image of surface deformation from FEA when cooling the Nickel-coated grating blank from room temperature to 80K. The PV value of 0.3 $\mu m$ is calculated for a square of 60mm x 60mm in the center of the grating. The deviations at the edges are larger.}
\label{fig:FEM}
\end{center}
\end{figure}

The 4 diffraction gratings in SPIFFI were direct ruled on identical blanks. The left panel of Figure \ref{fig:FEM} shows a CAD model of one of these blanks, which is 160mm x 140mm wide and 20mm thick with seven 15mm diameter lightweighting holes drilled through the grating in each direction. The blank is made of 6061 Aluminum, and was galvanically coated on all surfaces with $\sim$125 $\mu m$ layer of Nickel. The two large faces of the grating blank were then polished to be flat, and thus have $\sim$100 $\mu m$ of Nickel remaining. Finally, a thick layer of gold was applied to each side of the blank before the grating was ruled. We performed a Finite Element Analysis (FEA) of the grating blank with Nickel coating to determine the surface deformation when the grating is cooled from room temperature to 80K, the operating temperature of SPIFFI. The resulting surface deformation as a result of bimetallic bending stress is shown in the right panel of figure \ref{fig:FEM}. The deformation is a regular ``grid" pattern with a a PV deformation in the center of $\sim$300 nm, with larger deviations at the edges of the diffraction grating.

We set up a simple optical model of the spectrograph using each of the three J-K band gratings, and applied the FEA calculated surface deviation from figure \ref{fig:FEM} to the grating surfaces in the optical model. Additionally, we applied the measured collimator wavefront from figure \ref{fig:collimator} to the model collimator. We used the 1D partial coherence function in CodeV to simulate the effects of the different beam footprints in different pixel scales on the grating. We found that by scaling the amplitude of the FEA-calculated surface deviation of the grating by a factor 1.25-1.5, we were able to reproduce the line profile shapes we see in the different wavelengths in J-K bands. The ``grid" pattern of surface deviation acts like a second diffraction grating, creating shoulders on the main peak that start out very short at long wavelengths, and get higher as we move to shorter wavelengths. The separation of the shoulders from the main peak is determined only by the frequency of the grid (fixed by the position of the lightweighting holes), while the amplitude of the shoulders as a function of wavelength is determined by the total amplitude of the grid deformation. Since the grid is not perfectly sinusoidal, as we move to larger beam footprints, the line profiles become ``washed out" and line profile looks only broadened. This matches the behaviour seen in the instrument nearly perfectly. 

Since all the gratings were produced on nearly identical blanks (and thus should have very similar bimetallic surface deviations), this model allowed us to verify that all of the diffraction gratings should produce similar line profiles at a single wavelength. Note that we do not expect identical lines, as each grating is still unique--each grating has a different grating constant, overall surface form error, distance to the focal plane, and nickel layer thickness, and given the geometry of the instrument, is operating in a different diffraction order at a slightly different incidence angle for a single wavelength. However, the morphology of a line should be very similar between gratings. Additionally, while the model of bimetallic bending stress distorting the grating surface predicts the behaviour we see, we need to verify the actual distortion of the diffraction grating surface when cooled to liquid nitrogen temperatures and compare it to the results of the FEA. We have obtained a spare grating from the original manufacturing run, and plan to perform a wavefront measurement at 80K in a special test chamber at MPE. It should be noted that each grating blank was polished separately, and thus the layer thicknesses are not identical between the gratings, so the cryogenic measurements will only be representative of the actual distortion of the gratings in SPIFFI, not identical. 

Given the results of our finite element analysis, optical modelling, and tests in different diffraction orders on the SPIFFI gratings, we conclude that the bimetallic bending stress in the grating blanks is most likely to be responsible for the shoulders in the SPIFFI line profiles. Replacing the gratings with ones manufactured on solid blanks should give performance close to the design spectral resolution. This will have the largest effect in J band, but will benefit all wavelengths.

%%%%%%%%%%%%%%%%%%%%%%%%%%%%%%%%%%%%%%%%%%%%%%%%%%%%%%%%%%%%%
\subsection{Throughput}
\label{sec:throughput}

\begin{figure}[htbp!]
\begin{center}
\resizebox{0.9\textwidth}{!}{
\includegraphics[width=1.0\textwidth]{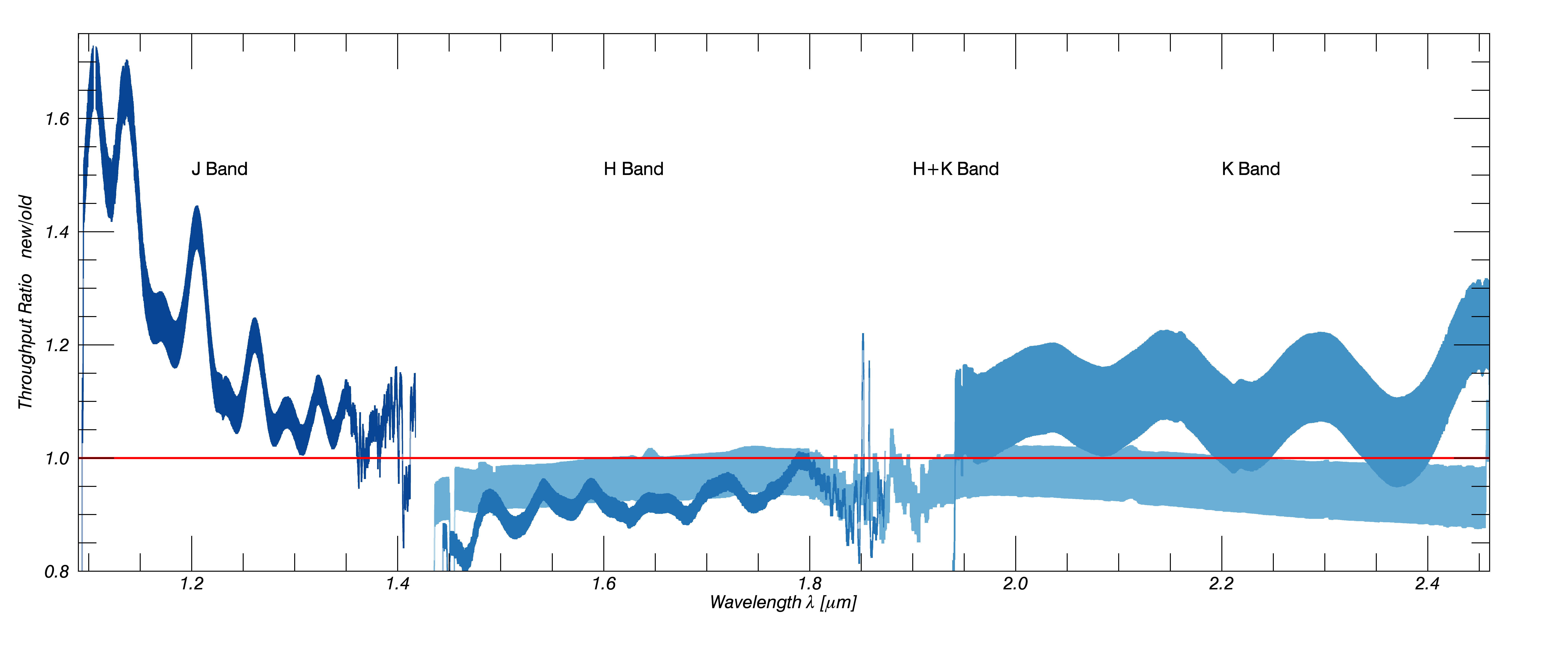}
}
\caption{Plot of the throughput ratio (new/old) over the wavelength range of the instrument. Each band is plotted separately, with the shaded regions indicating the standard deviation of the 4 ratios measured in each band. The shaded region does not include additional sources of systematic error such as dust buildup on the primary and secondary telescope mirrors during the two months between the measurements or potential pupil misalignments between SPIFFI and MACAO. These additional sources of error are estimated to be of order $\sim$10\%.}
\label{fig:throughput}
\end{center}
\end{figure}

A set of measurements on standard stars was taken both before and after the January 2016 upgrade in all bands and pixel scales in clear conditions. These stars were HD38921, HD49798, and HD75223. The first two stars were used to determine the throughput gain from the upgrade, as by the time the third star was measured, the seeing had deteriorated to $\gt$ 2". The flux from each star was integrated out to a radius on the detector determined by the point at which 99.9\% of the encircled energy was within that radius and corrected for airmass. This allows a partial compensation for variations in seeing and AO performance. The ratio of the flux before and after the upgrade as a function of wavelength was then calculated in each band. The ratios calculated from the 100 mas and 250 mas pixel scales and the two standard stars were averaged, and the result is plotted in figure \ref{fig:throughput}. The errors plotted only include the standard deviation of the 4 ratios in each band, and do not include additional effects such as dust buildup on the primary and secondary telescope mirrors during the two months between the measurements, potential pupil misalignments between SPIFFI and MACAO, or any other systematic errors. These additional sources of error are estimated to be of order $\sim$10\%. The throughput was substantially improved in the J and K bands, while remaining the same or even decreasing slightly in the H and H+K bands. More details on this measurement can be found in Gr\"aff 2016\cite{graeff16}.

%%%%%%%%%%%%%%%%%%%%%%%%%%%%%%%%%%%%%%%%%%%%%%%%%%%%%%%%%%%%%
\section{Planned upgrades for ERIS}
\label{sec:eris}

ERIS is the next-generation AO near-IR imager and spectrograph for the Cassegrain focus of the VLT UT4, and will use the Adaptive Optics Facility (AOF).\cite{amico12, kuntschner14} The ERIS instrument is made up of an adaptive optics module (ERIS-AO), a J-M band diffraction limited imager (NIX), a calibration unit (CU) and SPIFFIER. We briefly give an update of the ERIS project as it currently stands.

Since the publication of Kuntschner et al 2014\cite{kuntschner14}, the project has undergone significant design and consortium changes. The consortium has grown, and broadly speaking the new breakdown of responsibility is as follows: the Max-Planck Institut f\"ur Extraterrestrische Physik (MPE) is the PI insitute, responsible for the overall ERIS instrument as well as the upgrade of SPIFFI to SPIFFIER. UK Astronomy Techology center (UK-ATC) is responsible for the NIX camera in collaboration with Eidgen\"ossische Technische Hochschule Z\"urich (ETH) and NOVA. Osservatorio Astrofisico di Arcetri is responsible for the AO module, while Osservatorio Astronomico di Teramo is responsible for the calibration unit, and Osservatorio Astronomico di Padua is responsible for the instrument control software. ESO is responsible for the detectors and interface of ERIS with the VLT. 

\begin{figure}[htbp!]
\begin{center}
\resizebox{0.9\textwidth}{!}{
\includegraphics[trim={0 1.8cm 0 1.8cm}, clip=true, width=1.0\textwidth]{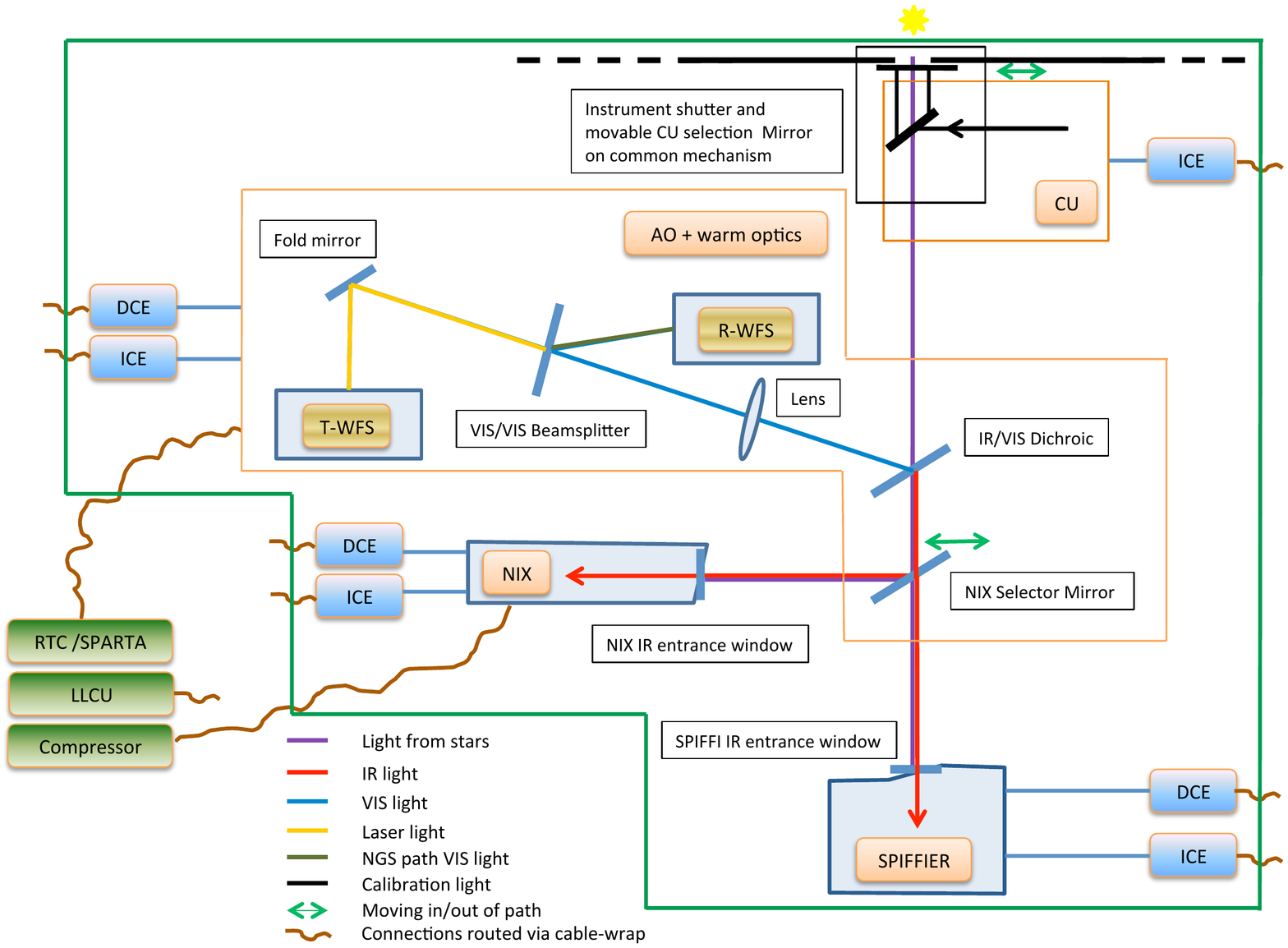}
}
\caption{Schematic of full ERIS optical layout. The beam path is as follows: star light (purple line) enters the instrument, goes through a warm dichroic beam splitter, and the optical light (blue line) goes to the AO system, while the IR light (red line) goes either into SPIFFI, or a pickoff mirror is moved into the beam to direct the light into NIX. The calibration unit light can be injected into the beam via another pickoff mirror.}
\label{fig:eris}
\end{center}
\end{figure}

The overall design of the instrument has become much simpler compared to the design presented in Kuntschner et al 2014\cite{kuntschner14}. There is no longer an IR wave front sensor (WFS), and the entrance windows of SPIFFIER and NIX are no longer dichroic, vastly simplifying the optical paths and alignment of the different subsystems. The beam path is as follows: light enters the instrument, goes through a warm dichroic beam splitter, and the optical light goes to the AO system, while the IR light goes either into SPIFFIER, or a pickoff mirror is moved into the beam to direct the light into NIX. The calibration unit light can be injected into the beam via another pickoff mirror. The AO system features two wavefront sensors, one for the LGS and one for the NGS, and will operate using the deformable mirror of the AOF. A detailed description of the new AO system design is provided elsewhere in these proceedings.\cite{riccardi16}. The NIX camera will provide J-M band imaging, coronography, sparse-aperture masking, and long-slit spectroscopy. A detailed design of the NIX camera is provided elsewhere in these proceedings.\cite{taylor16} The calibration unit will provide flat fields, wavelength calibration, and  distortion calibration in J-K bands for SPIFFIER and NIX, as well as calibrations for the AO system. The calibration unit design is available elsewhere in these proceedings.\cite{dolci16} Figure \ref{fig:eris} gives a schematic layout of the new instrument design with the different optical paths labeled.

%%%%%%%%%%%%%%%%%%%%%%%%%%%%%%%%%%%%%%%%%%%%%%%%%%%%%%%%%%%%%
\subsection{Diffraction gratings}
\label{sec:gratings}

In SPIFFIER, the H+K band grating currently installed in SPIFFI will be replaced with a high resolution grating providing R $\sim$ 8000 in J, H, and K bands, operating in 5th, 4th, and 3rd order respectively. A single exposure will cover half of one band, with 3 overlapping choices of wavelength coverage available in each band: the short-wavelength half, middle, or long-wavelength half of the band. The new wavelength ranges will also require new order sorting filters, which can be placed in the empty slots in the filter wheel. 

Given the geometry constraints in the SPIFFIER cryostat, a normal reflection grating is not the ideal choice for the high resolution grating, as there would be some vignetting in the 250mas pixel scale, and the maximum theoretical efficiency is $\sim60 \%$. Therefore, in 2014 we began an investigation into immersion grating development, which provides higher resolution in a smaller volume as well as higher efficiency (up to $\sim$80\%). We recently had a small (20mm x 20mm) ZnS prototype of immersion grating made to test whether an immersion grating would be feasible for SPIFFIER. AMOS developed the cutting technique, cut the ZnS prism, and measured the groove shapes. Fraunhofer IOF did development of the AR and Reflective coatings for ZnS, and we did cryogenic qualification of the coatings. The grating is currently at Fraunhofer IOF for coating and will return to AMOS in the next months for efficiency measurements. The decision of whether to use an immersion grating of reflection grating for the high-resolution grating will depend on the results of these measurements.

We also wish to improve the line profiles of the SPIFFIER instrument. If our hypothesis of cryogenic bimetallic bending stress in the gratings is correct (see section \ref{sec:lineexplain}), then replacing the current diffraction gratings with a set manufactured on blanks designed to have no bending stress would improve the line shape of the instrument to close to the design resolution. This is most relevant in J-band, however, the line shapes H and K bands would also benefit. The decision of whether or not to replace these gratings will be made after cryogenic wavefront measurements of a spare J-band grating are completed and the actual cold deformation of the grating surface is measured.

%%%%%%%%%%%%%%%%%%%%%%%%%%%%%%%%%%%%%%%%%%%%%%%%%%%%%%%%%%%%%
\subsection{Grating drive}
\label{sec:gratdrive}

Given the damage and wear to the grating drive and toothed wheel found after the 2016 upgrade, we are investigating a re-design of the grating drive. Currently the grating wheel is driven to the correct position, and a feedback loop between the encoder and motor keeps the wheel in place. This can result in``chattering" of the drive as the motor repeatedly switches between high and low power to keep the wheel's position, which could be an explanation for the excessive wear. One option we are considering would be to include an active break in the new grating drive that will be engaged once the grating wheel reaches it's desired position, allowing the drive motor to disengage.

%%%%%%%%%%%%%%%%%%%%%%%%%%%%%%%%%%%%%%%%%%%%%%%%%%%%%%%%%%%%%
\subsection{Detector and Persistence Mitigation}
\label{sec:detector}

\begin{figure}[htb!]
\begin{center}
\resizebox{0.9\textwidth}{!}{
\includegraphics[height=7cm]{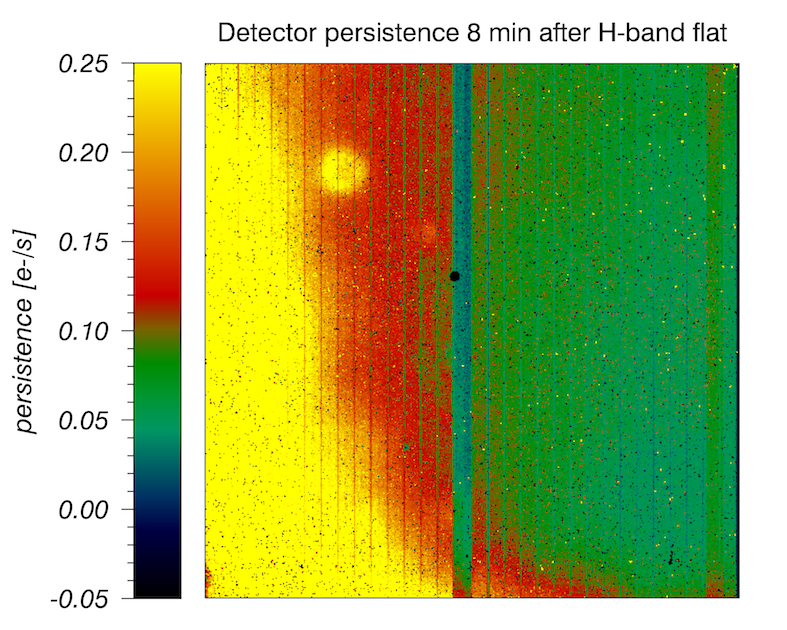}
\includegraphics[trim={4cm 0 0 0}, clip=true, height=7cm]{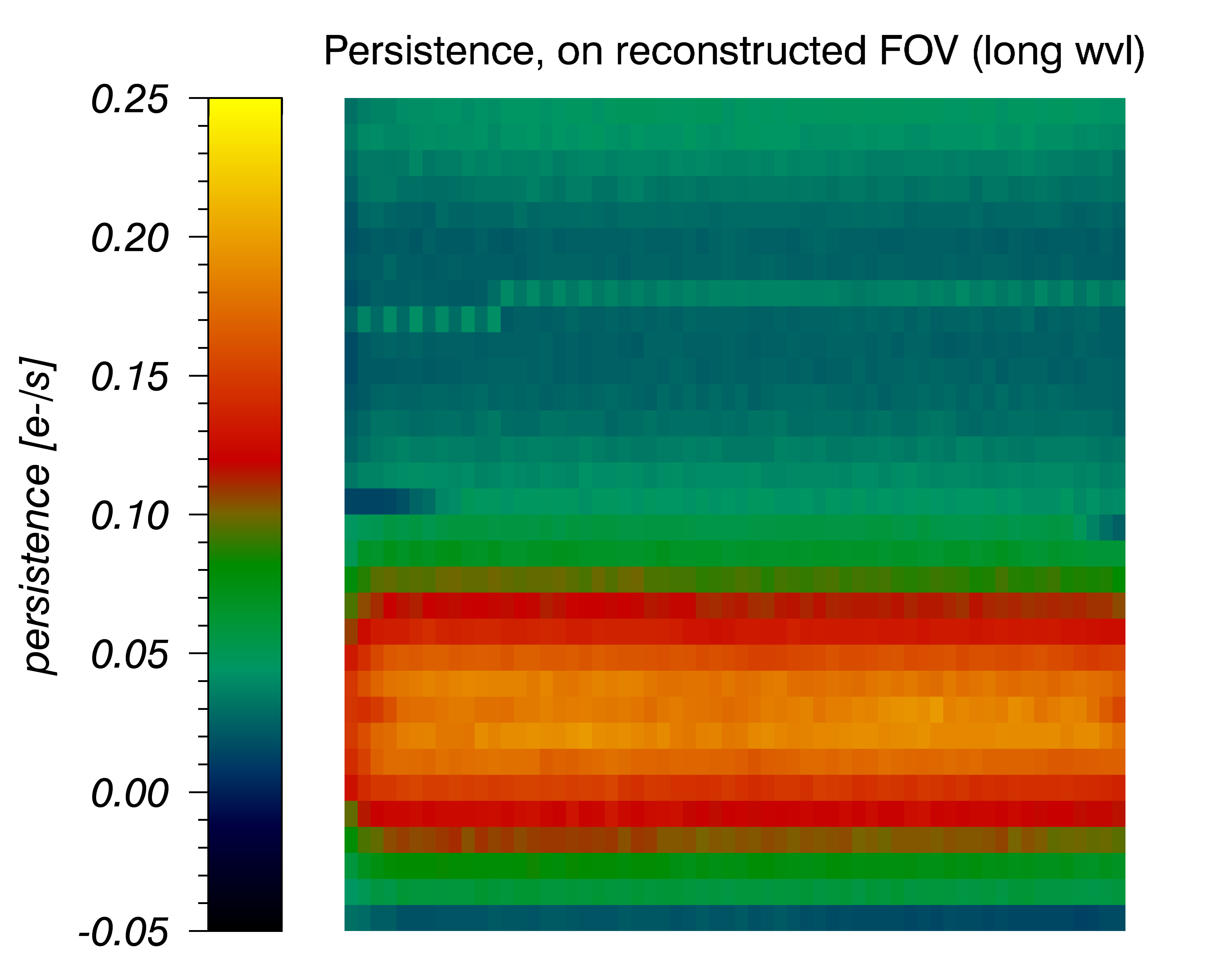}
}
\caption{{\it Left:} A 600s dark detector image taken after many repeated H-band flat field calibration frames, with peak illumination around 10,000 counts. The units are $e^{-}/s$, and the max value displayed is 0.25 (anything yellow is 0.25 $e^{-}/s$ of higher). There is a strong gradient in persistence across the detector. On the detector image, long wavelengths are at the bottom and short wavelengths are at the top. In general, small slitlet numbers are to the left, and high slitlet numbers are to the right, though the ordering of the slitlets is not sequential on the detector. {\it Right:} Reconstructed FOV with the image produced from the long-wavelength portion of the spectrum. Slitlet 1 is at the bottom, and slitlet 32 is at the top. The result of the persistence is the red stripe at the bottom of the detector.}
\label{fig:persistence}
\end{center}
\end{figure}

We plan to replace the current SPIFFI detector with a new HAWAII 2RG detector before integration into ERIS, as well as the replacement the IRACE detector control electronics with the New General Controller (NGC electronics). The current detector in SPIFFI is one of the very first science-grade HAWAII 2RG detectors ever made, and while it has served the instrument well for the past 11 years, it could be improved. It has a large circle of dead pixels in the center of the detector, as well as a lower quantum efficiency (QE) than is possible with the new HAWAII 2RG devices. There is also a strong gradient in persistence across the detector array, which when the detector image is reconstructed into a a 3D image cube, results in a strip of persistence in the lower half of the FOV. Figure \ref{fig:persistence} shows a detector dark image affected by persistence, as well as the reconstructed data cube. The detector replacement should improve QE, cosmetics, and persistence effects.

Science data is negatively affected by the persistence effect, and this region is often masked out in scientific analysis (see for example, figure 5 in Neumayer et al 2007\cite{neumayer07}). Until now the origin of this bright strip in the data cubes has been unclear to scientific users of SINFONI, and has been attributed to everything from illumination effects to the data reduction pipeline. Effort has gone into trying to empirically characterize and eliminate the persistence effect on the data (see, for example Menezes et al 2015\cite{menezes15} figures 14-17). 

We have conducted some short tests on the current SPIFFI detector and have been able to partially characterize the persistence decay and successfully subtract the effects of persistence from subsequent detector images at the $\sim95\%$ level based on a simple model of the detector persistence. We are currently working with the ESO detector group to develop a strategy for fully characterizing the the persistence behaviour of the new NIX and SPIFFIER detectors. Once the detector persistence behaviour is fully characterized, we plan to implement a persistence correction algorithm into the SPIFFIER and NIX pipelines similar to the one currently available for the Hubble Space Telescope Wide Field Camera 3 (HST WFC3).\cite{long12, mackenty14} 

%%%%%%%%%%%%%%%%%%%%%%%%%%%%%%%%%%%%%%%%%%%%%%%%%%%%%%%%%%%%%
\subsection{Motors and Electronics}
\label{sec:motors}

The standards for motor control and electronics at VLT has been moving towards industry standards, in particular, Programable Logic Controllers (PLCs) with Beckoff motion control.\cite{kiekebusch10, popovic14} Currently all of the motors in SPIFFI are 5-phase stepper motors. These motors will be replaced with 2-phase Phytron cryogenic stepper motors with Beckoff/PLC control. We have tested the new motors and control electronics cryogenically during the pre-optics test (driving the filter and pre-optics wheels), as well as on the spare grating drive, and they perform adequately. The cryo/vacuum controls will also be switched to PLC control.

%%%%%%%%%%%%%%%%%%%%%%%%%%%%%%%%%%%%%%%%%%%%%%%%%%%%%%%%%%%%%
\subsection{Other changes}
\label{sec:other}

There are a few other changes required to SPIFFI before integration into ERIS. The sky spider will be removed, as it is not forseen to be used in ERIS observations. The dichroic entrance window will be replaced with a normal entrance window. The cryostat lid will become thinner to allow more space for the ERIS warm optics. As a consequence, it will flex more during cryostat evacuation, and so the support structure for the metrology system must be re-designed. A cold finger will extend up from the SPIFFIER cold structure to illuminate the backside of the warm dichroic in ERIS. This will limit thermal background entering the SPIFFIER and NIX entrance windows. As a result, a vacuum tower + window assembly will be added to the cryostat lid. The pre-optics collimator will be replaced with one adapted to the new back focal length of the telescope of 500mm. These items are currently in their final design phase.

%%%%%%%%%%%%%%%%%%%%%%%%%%%%%%%%%%%%%%%%%%%%%%%%%%%%%%%%%%%%%
\section{Conclusion}
\label{sec:conclusion}

The SPIFFI instrument as part of SINFONI has been very scientifically productive, and is a key instrument for several high-profile science cases, including studying the galactic center and high redshift galaxy evolution. We are upgrading the SPIFFI instrument so it may be re-used as the subsystem SPIFFIER in ERIS. SPIFFIER will take advantage of improved performance of the the ERIS AO and the AOF. We completed the first half of the instrument upgrade in January 2016, and measured improvement in throughput, spectral, and spatial resolution of the SPIFFI instrument. The remaining items in the upgrade will be installed in late 2019, and will further improve the performance of the instrument, including the spectral resolution/line profiles of the instrument. The enhanced performance will benefit all science cases, increasing the scientific output of the instrument.

%%%%%%%%%%%%%%%%%%%%%%%%%%%%%%%%%%%%%%%%%%%%%%%%%%%%%%%%%%%%%
%%%%% References %%%%%

\bibliography{mpe}   %>>>> bibliography data in report.bib
\bibliographystyle{spiebib}   %>>>> makes bibtex use spiebib.bst

\end{document}